\documentclass[aps,pra,floatfix,a12]{revtex4-2}
\usepackage{amsfonts}
\usepackage{amsmath}
\usepackage{amsthm}
\usepackage{amssymb}
\usepackage{graphicx}
\usepackage{appendix}
\usepackage{hyperref}
\usepackage{textcomp}
\usepackage{multirow}
\usepackage{color}
\usepackage{bm}
\usepackage{braket}
\usepackage{subfigure}

\begin{document}

\title{Basic fractional nonlinear-wave models and solitons}
\author{Boris A. Malomed}
\affiliation{Department of Physical Electronics, School of Electrical Engineering, 
Faculty of Engineering, Tel Aviv University, Tel Aviv 69978, Israel\\
Instituto de Alta Investigaci\'{o}n, Universidad de Tarapac\'{a}, Casilla
7D, Arica, Chile}

\begin{abstract}
This review article provides a concise summary of one- and two-dimensional
models for the propagation of linear and nonlinear waves in fractional
media. The basic models, which originate from the Laskin's fractional
quantum mechanics and more experimentally relevant setups emulating
fractional diffraction in optics, are based on the Riesz definition of
fractional derivatives, which are characterized by the respective \textit{L%
\'{e}vy indices.} Basic species of one-dimensional solitons, produced by the
fractional models which include cubic or quadratic nonlinear terms, are
outlined too. In particular, it is demonstrated that the variational
approximation is relevant in many cases. A summary of the recently
demonstrated experimental realization of the fractional group-velocity
dispersion in fiber lasers is also presented.

\textbf{The paper is devoted to the celebration of the 80th birthday of
David K. Campbell.}
\end{abstract}

\maketitle


\noindent \textbf{The formal calculus based on the concept of fractional
derivatives is a branch of formal mathematics that has been known in the
course of ca. 200 years. Ca. 20 years ago, this concept had appeared in
physics as an ingredient of fractional quantum mechanics, which is based on
the Schr\"{o}dinger equation with the fractional operator of kinetic energy,
for the wave function of particles which, in their classical form, move by
random \textit{L\'{e}vy flights}. The fractionality of the Schr\"{o}dinger
equations of this type is characterized by the \textit{L\'{e}vy index} (LI) }%
$\alpha $\textbf{, which takes values }$0<\alpha \leq 2$\textbf{. It
determines the form of the fractional kinetic-energy operator, as }$\left(
-\partial ^{2}/\partial x^{2}\right) ^{\alpha /2}$\textbf{, the usual
(nonfractional) quantum mechanics corresponding to }$\alpha =2$\textbf{.
While fractional quantum mechanics remains far from experimental
implementation, much interest has been drawn to the more recent proposal to
emulate the fractional diffraction, modeled by the same equations, by means
of the wave propagation in specially devised optical cavities. Many
theoretical results have been reported for solitons, vortices and other
modes supported by the respective fractional Schr\"{o}dinger equations which
include cubic or quadratic optical nonlinearities. Very recently, the first
experimental implementation of the effective fractional group-velocity
dispersion has been reported for the light propagation in a fiber-laser
cavity. The great deal of the current interest to these theoretical and
experimental studies suggest relevance of providing a summary of the state
of art in the area. Such a concise summary is offered by this article, which
outlines basic models of the fractional wave propagation, and a brief
overview of basic species of solitons produced by these models.}

\section{Introduction: the concept of fractional-order derivatives}

\label{sec:1}

Ordinary derivatives are the central concept of the classical calculus,
which has been the mathematical basis of modern science in the course of the
last 350 years. In addition to the commonly known definition of derivatives
of integer orders, mathematical curiosity suggested attempts to introduce
derivatives of fractional orders. For example, the Darrieus-Landau
instability of plane combustion fronts, propagating in a premixed gas, is
determined by the linear dispersion relation between the instability growth
rate $\gamma $ and wavenumber $k$ of one-dimensional (1D) perturbations
corrugating the flat combustion front,
\begin{equation}
\gamma =C|k|,  \label{Landau}
\end{equation}%
with a constant $C>0$ \cite{Darrieus}. In Eq. (\ref{Landau}, $|k|$ implies
that the same instability gain is produced by right- and left-traveling
perturbations. In the framework of the phenomenological model \cite%
{Zeldovich}, one can introduce a model equation for a real order parameter $%
u\left( x,t\right) $ which corresponds to the dispersion relation (\ref%
{Landau}),%
\begin{equation}
\frac{\partial u}{\partial t}=C\sqrt{-\frac{\partial ^{2}}{\partial x^{2}}}u,
\label{sqrt}
\end{equation}%
where $\sqrt{-\partial ^{2}/\partial x^{2}}$is an operator that gives rise
to factor $|k|$ in the Fourier space, where $k$ is the wavenumber conjugate
to the spatial coordinate $x$ running along the flat combustion front.
Similarly, 2D perturbations with wave vector $\mathbf{k}=\left(
k_{x},k_{y}\right) $ give rise to the instability dispersion relation
\begin{equation}
\gamma =C|\mathbf{k}|,  \label{vector}
\end{equation}%
which suggests to introduce the respective 2D phenomenological equation,%
\begin{equation}
\frac{\partial u}{\partial t}=C\sqrt{-\frac{\partial ^{2}}{\partial x^{2}}-%
\frac{\partial ^{2}}{\partial y^{2}}}u\equiv \sqrt{-\Delta }u,  \label{2D}
\end{equation}%
where $\Delta $ is the usual 2D Laplacian acting on coordinates $\left(
x,y\right) $ on the flat combustion front.

It is shown in detail below that, while there are different formal
definitions of fractional differential operators, such as the one $\sqrt{%
-\Delta }$ in Eq. (\ref{2D}), the definitions which are relevant for\ the
realization in physics are defined in terms of the action of the operators
in the Fourier space, similar to the relation between Eqs. (\ref{sqrt}) and (%
\ref{Landau}) in 1D, or between Eqs. (\ref{2D}) and (\ref{vector}) in 2D.

Much earlier than fractional derivatives have appeared in physical models,
they were considered as a formal generalization of the classical calculus.
In a systematic form, the concept of the fractional calculus was elaborated
by Niels Henrik Abel in 1823 \cite{Abel} and Joseph Liouville in 1832 \cite%
{Liouville}. The development of these concepts has led to the abstract
definition of what is known as the \textit{Caputo derivative} of non-integer
order $\alpha $ \cite{Uchaikin,Caputo},%
\begin{equation}
D_{x}^{\alpha }\psi (x)=\frac{1}{\Gamma \left( 1-\{\alpha \}\right) }%
\int_{0}^{x}\frac{f^{\left( n\right) }(\xi )dx}{\left( x-\xi \right)
^{\left\{ \alpha \right\} }},  \label{cap}
\end{equation}%
where $n\equiv \lbrack \alpha ]+1$, $[\alpha ]$ stands for the integer part
of the fractional order, $\{\alpha \}\equiv \alpha -[\alpha ]$ is its
fractional part, $\Gamma $ is the Gamma-function, and $f^{(n)}$ is the usual
derivative. Note that this definition represents the fractional derivative
as integral (nonlocal) operator, rather than as a differential one. As shown
below, the fractional derivatives which appear in models originating in
physics are defined differently, namely, as \textit{Riesz derivatives} \cite%
{Riesz}, see Eqs. (\ref{Riesz derivative}) and (\ref{2D operator}) in
Section 2.

This paper aims to produce a concise overview of linear and nonlinear
dynamics in physically relevant models of fractional media. Also included is
a summary of the first experimental realization of the concept of the
fractional derivatives, in the form of a fiber cavity which emulates
fractional group-velocity dispersion (GVD) \cite{Shilong}. The presentation
is not comprehensive; in particular, while the main objective is to
introduce models of linear and nonlinear fractional media which are relevant
to physics, the summary of soliton solutions amounts to the presentation of
a few examples, many other cases being only briefly mentioned. An earlier
review of theoretical results for solitons in fractional models is provided
in Ref. \cite{review}.

\section{The advent of fractional calculus to physics: Fractional quantum
mechanics}

\label{sec:2}

For the first time, fractional derivatives had appeared in the context of
physically relevant models as the mathematical basis of\ \textit{fractional
quantum mechanics}. This theory was introduced by Laskin \cite{Lask1,Lask2}
for nonrelativistic particles which move, at the classical level, by \textit{%
L\'{e}vy flights }(random leaps).

\subsection{Classical particles moving by L\'{e}vy flights}

The starting point for the development of fractional quantum mechanics is to
consider particles whose classical stochastic motion is performed as a chain
of random leaps (\textquotedblleft \textit{L\'{e}vy flights}"), rather than
in the form of the usual Brownian random walk. In this regime, the mean
distance $|x|$ of the particle, which moves by 1D \textquotedblleft flights"
from the initial position, $x=0$, grows with time $t$ as%
\begin{equation}
|x|\sim t^{1/\alpha },  \label{flight}
\end{equation}%
where the \textit{L\'{e}vy index} (LI) $\alpha $ takes values%
\begin{equation}
0<\alpha \leq 2  \label{LI}
\end{equation}%
\cite{Mandelbrot}. The limit value, $\alpha =2$, corresponds to the usual
Brownian random walk, while at $\alpha <2$ Eq. (\ref{flight}) demonstrates
that the L\'{e}vy flights give rise to faster growth of $|x|$ at $%
t\rightarrow \infty $. As concerns realizations of the L\'{e}vy flights in
nature, a usual reference is to the life style of sharks in the ocean, which
search for food moving by random fast swims, even if sharks are not
appropriate objects for considering their motion in the quantum regime,
which is the next step of the derivation, leading to the \textit{fractional
Schr\"{o}dinger equation} (FSE)

\subsection{The fractional linear Schr\"{o}dinger equations for L\'{e}%
vy-flying particles and the fractional Riesz derivative}

The derivation of an effective Schr\"{o}dinger equation as a result of the
quantization of a particle which moves, at the classical level, by L\'{e}vy
flight was elaborated by Laskin \cite{Lask1,Lask2} (see also Ref. \cite%
{GuoXu}). It is based on the fundamental formalism which defines\ quantum
mechanics by means of the Feynman's path integration. This formalism
represents the quantum dynamics of a particle as a result of the
superposition of virtual motions along all randomly chosen trajectories
(paths). The superposition is defined as the integral in the space of all
paths, $\sim \int \exp \left[ iS(\mathrm{path})\right] d(\mathrm{path})$,
where $S$ is the classical action corresponding to a particular path \cite%
{Feynman}. To apply this concept, Laskin considered the superposition of the
paths which correspond not to the Brownian random walks, but to the L\'{e}vy
flights. In the 1D setting, the result is FSE for wave function $\Psi \left(
x,t\right) $. In the scaled form, it is written as
\begin{equation}
i\frac{\partial \Psi }{\partial t}=\frac{1}{2}\left( -\frac{\partial ^{2}}{%
\partial x^{2}}\right) ^{\alpha /2}\Psi +V(x)\Psi ,  \label{FSE}
\end{equation}%
where $V(x)$ is the same external potential which appears in the usual Schr%
\"{o}dinger equation.

A fundamentally important result of the derivation of Eq. (\ref{FSE}) by
means of the path-integral formalism is that the usual kinetic-energy
operator, $-(1/2)\partial ^{2}/\partial x^{2}$, which corresponds to $\alpha
=2$ in Eq. (\ref{FSE}), is replaced, at $\alpha <2$, by the fractional
\textit{Riesz derivative} \cite{Riesz}, which is defined as the
juxtaposition of the direct and inverse Fourier transforms ($x\rightarrow
p\rightarrow x$) of the wave function,
\begin{equation}
\left( -\frac{\partial ^{2}}{\partial x^{2}}\right) ^{\alpha /2}\Psi =\frac{1%
}{2\pi }\int_{-\infty }^{+\infty }dp|p|^{\alpha }\int_{-\infty }^{+\infty
}d\xi e^{ip(x-\xi )}\Psi (\xi ).  \label{Riesz derivative}
\end{equation}%
This definition implies that the action of the fractional derivative, $%
\left( -\partial ^{2}/\partial x^{2}\right) ^{\alpha /2}$, in the Fourier
space amounts to the straightforward multiplication: $\hat{\Psi}%
(p)\rightarrow |p|^{\alpha }\hat{\Psi}(p)$, where
\begin{equation}
\hat{\Psi}(p)=\int_{-\infty }^{+\infty }\exp \left( -ipx\right) \Psi (x)dx
\label{Fourier}
\end{equation}%
is the Fourier transform of the wave function. In definition (\ref{Riesz
derivative}), $\alpha $ is the same LI which determines the classical L\'{e}%
vy flights as per Eq. (\ref{flight}).

Thus, the derivation of the physically relevant model, \textit{viz}., FSE,
leads to the relatively simple Riesz fractional derivative. As concerns
\textquotedblleft more sophisticated" varieties of fractional derivatives,
defined on abstract mathematical grounds, such as the Caputo derivative (\ref%
{cap}), they do not emerge naturally in these physical contexts.

In the 2D space, the same derivation gives rise to the 2D FSE for wave
function $\Psi \left( x,y,t\right) $) \cite{Lask2}:
\begin{equation}
i\frac{\partial \Psi }{\partial t}=\frac{1}{2}\left( -\frac{\partial ^{2}}{%
\partial x^{2}}-\frac{\partial ^{2}}{\partial y^{2}}\right) ^{\alpha /2}\Psi
+V(x,y)\Psi .  \label{2D FSE}
\end{equation}%
The fractional operator of the 2D kinetic energy is represented, in the 2D
Fourier space $\left( p,q\right) $, by multiplier $\left( p^{2}+q^{2}\right)
^{\alpha /2}$, i.e., its action amounts to
\begin{equation}
\hat{\Psi}(p,q)\rightarrow \left( p^{2}+q^{2}\right) ^{\alpha /2}\hat{\Psi}%
(p,q),  \label{Fourier 2D}
\end{equation}
where the 2D\ Fourier transform of the wave function
\begin{equation}
\hat{\Psi}(p,q)=\int \int \exp \left( -ipx-iqy\right) \Psi (x,y)dxdy,
\label{2D Fourier}
\end{equation}
cf. Eq. (\ref{Fourier}). Therefore, the action of the 2D operator in the
coordinate space is defined as the juxtaposition of the direct and inverse
Fourier transforms, with the multiplication as per Eq. (\ref{Fourier 2D})
inserted between them:
\begin{equation}
\left( -\frac{\partial ^{2}}{\partial x^{2}}-\frac{\partial ^{2}}{\partial
y^{2}}\right) ^{\alpha /2}\Psi =\frac{1}{(2\pi )^{2}}\int \int dpdq\left(
p^{2}+q^{2}\right) ^{\alpha /2}\int \int d\xi d\eta e^{i\left[ p(x-\xi
)+iq(y-\eta )\right] }\Psi (\xi ,\eta ),  \label{2D operator}
\end{equation}%
cf. the 1D equation (\ref{Riesz derivative}).

Stationary solutions to Eq. (\ref{2D FSE}), with real energy eigenvalue $\mu
$ (in terms of Bose-Einstein condensates (BECs), it is the \textit{chemical
potential}) are looked for in the usual form,%
\begin{equation}
\Psi \left( x,y,t\right) =e^{-i\mu t}U\left( x,y\right) ,  \label{U}
\end{equation}%
with real function $U$ satisfying the stationary equation,%
\begin{equation}
\mu U=\frac{1}{2}\left( -\frac{\partial ^{2}}{\partial x^{2}}-\frac{\partial
^{2}}{\partial y^{2}}\right) ^{\alpha /2}U+V(x,y)U,  \label{Ustat}
\end{equation}%
or its 1D version, without coordinate $y$. It is relevant to mention that in
the case of potential in the form of the harmonic oscillator, $V\left(
x,y\right) =\left( \Omega ^{2}/2\right) \left( x^{2}+y^{2}\right) $, the
Fourier transform (\ref{2D Fourier}) satisfies the Fourier transform of Eq. (%
\ref{2D FSE}), which takes the form of the usual Schr\"{o}dinger equation,
but written in the Fourier space \cite{nonlocal}:%
\begin{equation}
i\frac{\partial \hat{\Psi}}{\partial t}=-\frac{\Omega ^{2}}{2}\left( \frac{%
\partial ^{2}}{\partial p^{2}}+\frac{\partial ^{2}}{\partial q^{2}}\right)
\hat{\Psi}+\frac{1}{2}\left( p^{2}+q^{2}\right) ^{\alpha /2}\hat{\Psi},
\label{in-Fourier}
\end{equation}%
with the effective potential $(1/2)\left( p^{2}+q^{2}\right) ^{\alpha /2}$,
or the 1D version of this equation, in the absence of variable $q$. In
particular, a solution of the 1D version of Eq. (\ref{in-Fourier}) with $%
\alpha =1$, i.e., with the effective Fourier-space potential $(1/2)|p|$, can
be expressed in terms of the Airy function \cite{nonlocal}.

The fractional calculus has found another important application to physics in the
form of fractional kinetics. This concept, which was elaborated by Zaslavsky \textit{et al}.
\cite{Za1,Za2}, addresses, in particular, the kinetic theory, based on the
fractional Fokker-Planck equations, for dynamics which may be intermediate between
integrable and purely chaotic. It predicts fundamental effects such as anomalous
transport and superdiffusion, and models real kinetic phenomena such as advection
of particles in diverse physical settings. However, this topic is not considered in
detail in the present article.

\subsection{A conjecture: fractional Gross-Pitaevskii equations (FGPEs)}

It is commonly known that the dynamics of BECs in ultracold atomic gases is
very accurately approximated, in the framework of the mean-field theory, by
the Gross-Pitaevskii equation, in the form of the Schr\"{o}dinger equation
for single-particle wave function $\psi $, supplemented by the cubic terms
which represent the effect of inter-particle collisions \cite{Pit-Str}:%
\begin{equation}
i\hbar \frac{\partial \psi }{\partial t}=-\frac{\hbar ^{2}}{2m}\nabla
^{2}\psi +V(\mathbf{r})\psi +\frac{4\pi \hbar ^{2}}{m}a_{s}|\psi |^{2}\psi .
\label{GP}
\end{equation}%
Here $m$ is the atomic mass, and $a_{s}$ is the scattering length which
determines two-particle collisions ($a_{s}>0$ and $a_{s}<0$ correspond to
repulsive and attractive interactions between the particles, respectively).
An intriguing possibility is to consider a gas of bosonic particles obeying
the FSE (\ref{FSE}) or its 2D version (\ref{2D FSE}). While a systematic
derivation of the respective\textit{\ fractional Gross-Pitaevskii equation}
(FGPE), with the usual kinetic-energy operator in Eq. (\ref{GP}) replaced by
its fractional counterpart defined as per Eqs. (\ref{FSE}) or (\ref{2D FSE}%
), remains a challenging objective, a natural expectation is that the
equation sought for will take the following form, in the scaled form:%
\begin{equation}
i\frac{\partial \Psi }{\partial t}=\frac{1}{2}\left( -\frac{\partial ^{2}}{%
\partial x^{2}}-\frac{\partial ^{2}}{\partial y^{2}}\right) ^{\alpha /2}\Psi
+V(x)\Psi +\sigma |\Psi |^{2}\Psi ,  \label{FGPE}
\end{equation}%
where $\sigma =+1$ or $-1$ corresponds to the self-repulsive or attractive
condensate, respectively \cite{review}.

A recently elaborated example of the FGPE system is a model of a binary
(two-component) fractional BEC, described by a spinor wave function with
components $\left( \phi _{+},\phi _{-}\right) $, which maintains the \textit{%
spin-orbit coupling} (SOC) between the components \cite{SOC}. The respective
system of coupled FGPEs in 2D is \cite{HS}%
\begin{eqnarray}
i\frac{\partial \phi _{+}}{\partial t} &=&\frac{1}{2}\left( -\frac{\partial
^{2}}{\partial x^{2}}-\frac{\partial ^{2}}{\partial y^{2}}\right) ^{\alpha
/2}\phi _{+}-(|\phi _{+}|^{2}+\gamma |\phi _{-}|^{2})\phi _{+}+\lambda
\left( \frac{\partial \phi _{-}}{\partial x}-i\frac{\partial \phi _{-}}{%
\partial y}\right) ,  \notag \\
&&  \label{+-} \\
i\frac{\partial \phi _{-}}{\partial t} &=&\frac{1}{2}\left( -\frac{\partial
^{2}}{\partial x^{2}}-\frac{\partial ^{2}}{\partial y^{2}}\right) ^{\alpha
/2}\phi _{-}-(|\phi _{-}|^{2}+\gamma |\phi _{+}|^{2})\phi _{-}-\lambda
\left( \frac{\partial \phi _{+}}{\partial x}+i\frac{\partial \phi _{+}}{%
\partial y}\right) ,  \notag
\end{eqnarray}%
with the kinetic-energy operator defined as per Eq. (\ref{2D operator}), cf.
Eq. (\ref{FGPE}). Here, the attractive sign of the nonlinear interactions is
assumed, to provide a possibility to create solitons. The coefficient of the
self-attraction is scaled to be $1$, while $\gamma >0$ is the relative
strength of the cross-attraction between the components, and $\lambda $ is
the strength of the linearly-mixing terms representing the SOC. By means of
additional scaling, it is possible to set $\lambda \equiv 1/2$, while
soliton states with real chemical potential $\mu $ are looked for as%
\begin{equation}
\phi _{\pm }=e^{-i\mu t}u_{\pm }(x,y)  \label{chem pot}
\end{equation}%
(cf. Eq. (\ref{U})), where complex stationary wave functions $u_{\pm }$
satisfy equations%
\begin{eqnarray}
\mu u_{+} &=&\frac{1}{2}\left( -\frac{\partial ^{2}}{\partial x^{2}}-\frac{%
\partial ^{2}}{\partial y^{2}}\right) ^{\alpha /2}u_{+}-(|u_{+}|^{2}+\gamma
|u_{-}|^{2})u_{+}+\frac{1}{2}\left( \frac{\partial u_{-}}{\partial x}-i\frac{%
\partial u_{-}}{\partial y}\right) ,  \notag \\
&&  \label{u2D} \\
\mu u_{-} &=&\frac{1}{2}\left( -\frac{\partial ^{2}}{\partial x^{2}}-\frac{%
\partial ^{2}}{\partial y^{2}}\right) ^{\alpha /2}u_{-}-(|u_{-}|^{2}+\gamma
|u_{+}|^{2})u_{-}-\frac{1}{2}\left( \frac{\partial u_{+}}{\partial x}+i\frac{%
\partial u_{+}}{\partial y}\right) .  \notag
\end{eqnarray}%
Soliton solutions are characterized by the total norm,%
\begin{equation}
N\equiv \int \int \left( \left\vert u_{+}\right\vert ^{2}+\left\vert
u_{-}\right\vert ^{2}\right) dxdy.  \label{N2D}
\end{equation}

It is relevant to mention that the stationary system (\ref{u2D}) can be
derived from the respective \textit{Lagrangian}:%
\begin{gather}
L=\mu \int_{-\infty }^{+\infty }dx\int_{-\infty }^{+\infty }dy\left[
\left\vert u_{+}(x,y)\right\vert ^{2}+\left\vert u_{-}(x,y)\right\vert ^{2}%
\right]  \notag \\
-\frac{1}{2\pi ^{2}}\int_{0}^{+\infty }dp\int_{0}^{+\infty
}dq(p^{2}+q^{2})^{\alpha /2}\int_{-\infty }^{+\infty }d\xi \int_{-\infty
}^{+\infty }d\eta \int_{-\infty }^{+\infty }dx\int_{-\infty }^{+\infty
}dy\cos \left[ p(x-\xi )+q(y-\eta )\right]  \notag \\
\times \left[ u_{+}^{\ast }(x,y)u_{+}(\xi ,\eta )+u_{-}^{\ast
}(x,y)u_{-}(\xi ,\eta )\right]  \notag \\
-\frac{1}{2}\int_{-\infty }^{+\infty }dx\int_{-\infty }^{+\infty }dy\left[
u_{+}^{\ast }\frac{\partial u_{-}}{\partial x}+u_{+}\frac{\partial
u_{-}^{\ast }}{\partial x}-i\left( u_{+}^{\ast }\frac{\partial u_{-}}{%
\partial y}-u_{+}\frac{\partial u_{-}^{\ast }}{\partial y}\right) \right]
\notag \\
+\int_{-\infty }^{+\infty }dx\int_{-\infty }^{+\infty }dy\left[ \frac{1}{2}%
\left( |u_{+}|^{4}+|u_{-}|^{4}\right) +\gamma |u_{+}|^{2}|u_{-}|^{2}\right] ,
\label{L2D}
\end{gather}%
where $\ast $ stand for the complex conjugate, and the integration with
respect to $p$ and $q$ is reduced from the original domains $\left( -\infty
,+\infty \right) $ to $\left( 0,\infty \right) $, using the domains'
symmetry. Although expression (\ref{L2D}) seems cumbersome, it can be used
to predict solitons solutions of Eq. (\ref{u2D}) by means of the variational
approximation (VA). To this end, the \textit{ansatz} for the approximate
solutions is adopted as
\begin{equation}
u_{+}=A_{+}\exp (-\beta (x^{2}+y^{2})),u_{-}=A_{-}(x+iy)\exp (-\beta
(x^{2}+y^{2})),\beta >0,  \label{ans2D}
\end{equation}%
with inverse width $1/\sqrt{\beta }$ and amplitudes $A_{\pm }$ (cf. the
known version of the VA for the usual 2D nonlinear Schr\"{o}dinger equation
\cite{Desaix,Dimitrevski}). The norm of ansatz (\ref{ans2D}) is $N=\left(
\pi /2\beta \right) \left( A_{+}^{2}+A_{-}^{2}/(2\beta )\right) $.

\begin{figure}[h]
\begin{center}
\includegraphics[height=5.cm]{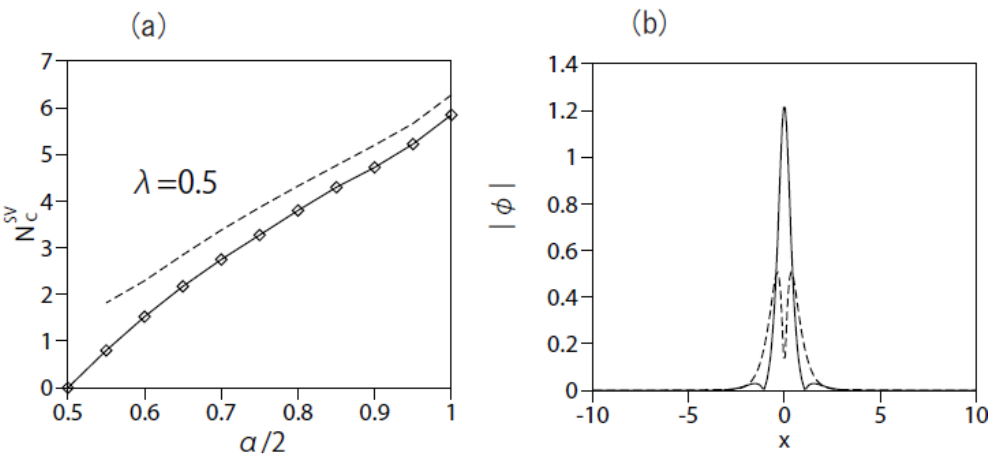}
\end{center}
\caption{(a) The 2D SV solitons exist and are stable at norms below the
critical value, $N_{\mathrm{c}}^{\mathrm{(SV)}}$, which is plotted vs. the
half-LI, $\protect\alpha /2$, for $\protect\lambda =0.5$ and $\protect\gamma %
=0$ in Eq. (\protect\ref{2D}). The collapse takes place at $N>N_{\mathrm{c}%
}^{\mathrm{(SV)}}$. The numerical result and its VA-predicted counterpart
are shown by the dashed line and chain of rhombuses, respectively. (b) Cross
sections of profiles of $\left\vert \protect\phi _{+}\right\vert $ and $%
\left\vert \protect\phi _{-}\right\vert $ in the SV state (continuous and
dashed lines, respectively) produced by the numerical solution of the
linearized version of Eq. (\protect\ref{u2D}) with $\protect\alpha =1$ (see
the text). Reproduced from H. Sakaguchi and B. A. Malomed, One- and two-dimensional
solitons in spin-orbit-coupled Bose-Einstein condensates with fractional
kinetic energy, J. Phys. B: At. Mol. Opt. Phys. \textbf{55}, 155301 (2022) [see Ref. \cite{HS}],
with the permission of IOP Publishing.}
\label{Fig1}
\end{figure}

The ansatz (\ref{ans2D}) represents the \textit{semi-vortex} (SV) type,
which implies that components $u_{+}$ and $u_{-}$ carry vorticities $0$ and $%
1$, respectively \cite{Ben Li} (see an example shown by means of
cross-sections of the two components in Fig. \ref{Fig1}(b)). The SV
structure is an exact feature of the 2D solitons shaped by SOC, which is not
restricted by the applicability of VA.

The substitution of ansatz (\ref{ans2D}) in Lagrangian (\ref{L2D}) produces
the respective \textit{effective Lagrangian},
\begin{equation}
L_{\mathrm{eff}}=\mu N-\frac{\pi \Gamma \left( 1+\alpha /2\right) }{2\left(
2\beta \right) ^{1-\alpha /2}}\left[ A_{+}^{2}+\left( 1+\frac{\alpha }{2}%
\right) \frac{A_{-}^{2}}{2\beta }\right] -\frac{\pi \lambda }{\beta }%
A_{+}A_{-}+\frac{\pi }{8\beta }\left( A_{+}^{4}+\frac{A_{-}^{4}}{8\beta ^{2}}%
+\frac{\gamma A_{+}^{2}A_{-}^{2}}{2\beta }\right) ,  \label{va1}
\end{equation}%
Then, for given $\mu $, SV's\ parameters are predicted by the Euler-Lagrange
equations, $\partial L_{\mathrm{eff}}/\partial A_{\pm }=\partial L_{\mathrm{%
eff}}/\partial \beta =0$.

Figure \ref{Fig1}(a) demonstrates that, as predicted by the VA and confirmed
by numerical results, the fractional SOC system gives rise to \emph{stable
SVs} in the interval of LI values
\begin{equation}
1<\alpha \leq 2,  \label{1-2}
\end{equation}
at norms below a critical value, $N\,<N_{\mathrm{c}}^{\mathrm{(SV)}}(\alpha
) $, while at $N\,>N_{\mathrm{c}}^{\mathrm{(SV)}}(\alpha )$ the solitons are
destroyed by the collapse, i.e., catastrophic self-compression of the field
leading to the appearance of a singularity after a finite evolution time.
The proximity of the VA-predicted and numerically found stability boundaries
in Fig. \ref{Fig1}(a) demonstrates that the VA, based on the simple ansatz (%
\ref{ans2D}), may produce reasonable results even for the complex system
including the fractional diffraction (kinetic-energy operator), SOC, and the
nonlinear interactions.

An exact property of the system is that $N_{\mathrm{c}}^{\mathrm{(SV)}%
}(\alpha =1)=0$, i.e., no SVs exist at $\alpha \leq 1$. In the case of $%
\alpha -1\rightarrow +0$, the SV exists with an infinitesimal amplitude,
hence it can be found as a numerical solution of the linearized version of
Eq. (\ref{u2D}) with $\alpha =1$. Cross-sections of components $\left\vert
u_{\pm }\right\vert $ of the latter solution are displayed in Fig. \ref{Fig1}%
(b).\ Actually, these plots adequately represent the shape of the SV
solitons in the general case, when the nonlinearity is present.

The SV solitons are stable for $\gamma <1$ in Eqs. (\ref{+-}). For $\gamma
>1 $, the SVs are unstable, while in this case there are stable solitons in
the form of \textit{mixed modes}, which mix zero-vorticity terms and ones
with vorticities $+1$ and $-1$, see further details in Ref. \cite{HS}.

\section{The Emulation of the fractional Schr\"{o}dinger equations (FSEs) in
optics}

\label{sec:3}

\subsection{Linear equations for the paraxial light propagation in optical
systems emulating fractional diffraction}

Realizations of the fractional quantum systems were proposed in solid-state
settings, such as Levy crystals \cite{Levy crystal} and exciton-polariton
condensates in semiconductor microcavities \cite{Pinsker}. However, no
experimental demonstration of the fractional quantum mechanics has been
demonstrated thus far.

\begin{figure}[h]
\begin{center}
\includegraphics[height=5.cm]{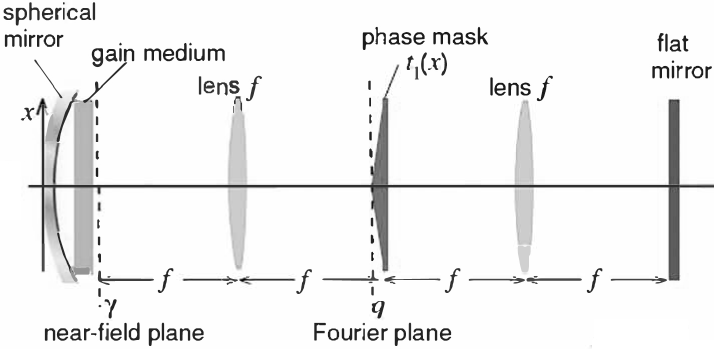}
\end{center}
\caption{The setup proposed in Ref. \protect\cite{Longhi} for the emulation
of the FSE in the optical cavity. The left and right lenses perform,
respectively, the spectral (Fourier) decomposition and recombination of the
light beam with respect to the transverse coordinates. The phase mask placed
in the central Fourier plane introduces the differential phase shift which
emulates the fractional diffraction,as per Eqs. (\protect\ref{R}) and (%
\protect\ref{Z}). The spherical mirror and gain medium at the left edge of
the cavity introduce an effective potential.
Reproduced with permission from
S. Longhi, Fractional Schr\"{o}dinger equation in optics,
Opt. Lett. \textbf{40}, 1117-1120 (2015)
[see Ref. \cite{Longhi}].
Copyright 2015, Optica.}
\label{Fig2}
\end{figure}

A more promising approach to the physical realization of FSEs in the form of
classical equations is suggested by the commonly known similarity of the Schr%
\"{o}dinger equation in quantum mechanics and the propagation equation for
the amplitude of the classical optical field in the usual case of paraxial
diffraction, with time $t$ replaced, as the evolution variable, by the
propagation distance, $z$ \cite{KA}. A scheme for the realization of this
possibility was proposed by Longhi in 2015 \cite{Longhi}, who considered the
transverse light dynamics in optical cavities with the $4f$
(four-focal-lengths) structure. As shown in Fig. \ref{Fig2}, the proposed
setup incorporates two lenses and a phase mask, which is placed in the
middle (Fourier) plane. The lenses perform the direct and inverse Fourier
transforms of the light beam with respect to the single or two transverse
coordinates, thus implementing Eq. (\ref{Fourier}) or (\ref{2D Fourier}) and
the inverse form of these equations. The Fourier decomposition splits the
beam into spectral components with transverse wavenumbers $\left( p,q\right)
$, the distance of each component from the optical axis being, roughly
speaking,
\begin{equation}
R\sim \sqrt{p^{2}+q^{2}}  \label{R}
\end{equation}

The action of the fractional diffraction in the Fourier space amounts,
according to Eq. (\ref{Riesz derivative}) or (\ref{2D operator}), to the
local phase shift
\begin{equation}
\hat{\Psi}\left( p,q\right) \rightarrow \hat{\Psi}\left( p,q\right) \exp %
\left[ i\left( p^{2}+q^{2}\right) ^{\alpha }Z\right] ,  \label{Z}
\end{equation}%
where $Z$ is the propagation distance which accounts for the fractional
diffraction. The corresponding differential phase shift of the Fourier
components is imposed by the phase mask in Fig. \ref{Fig2}, which is
designed so that the local shift introduced my the mask at distance $R$ from
the axis emulates relation (\ref{Z}), with $R$ and $\left( p,q\right) $
related according to Eq. (\ref{R}). In reality, the required mask, which is
the central element of the setup, can be created as a computer-generated
hologram \cite{Shilong}. Thus, the light beam recombined by the right lens
in Fig. \ref{Fig2} carries the phase structure corresponding to the action
of the fractional diffraction.

In addition to that, the curved mirror at the left edge of the cavity in
Fig. \ref{Fig2} introduces (prior to the action of the Fourier
decomposition) a phase shift which represents the action of potential $%
V\left( x,y\right) $ in Eq. (\ref{2D FSE}). The layer of the gain medium
placed next to the mirror in Fig. \ref{Fig2} may be used to amplify
particular modes in the transverse structure of the light beam.

The setup which is outlined above provides a single-step transformation of
the optical beam. The continuous FSE (\ref{2D FSE}) is then introduced as an
approximation for many cycles of circulation of light in the cavity,
assuming that each cycle introduces a small phase shift (\ref{Z}). The 1D
version of the FSE is obtained as the obvious reduction of the 2D scheme.

\subsection{The emulation of the fractional diffraction in nonlinear optical
systems: Fractional nonlinear Schr\"{o}dinger equations (FNLSEs) in the
spatial domain}

\subsubsection{The cubic nonlinearity in 1D}

Once the FSE may be implemented as the propagation equation in optics under
the action of the effective fractional diffraction, a natural possibility is
to include the self-focusing nonlinearity of the optical material in the
same system. The accordingly modified one-dimensional Eq. (\ref{FSE}) is the
fractional nonlinear Schr\"{o}dinger equation (FNLSE),
\begin{equation}
i\frac{\partial \Psi }{\partial z}=\frac{1}{2}\left( -\frac{\partial ^{2}}{%
\partial x^{2}}\right) ^{\alpha /2}\Psi +V(x)\Psi -g|\Psi |^{2}\Psi ,
\label{FNLSE}
\end{equation}%
where $g>0$ is the coefficient of the Kerr-nonlinearity coefficient, cf.
FGPE (\ref{FGPE}). Then, steady-state solutions to Eq. (\ref{FNLSE}), with
real propagation constant $k$, are looked for as
\begin{equation}
\Psi (x,z)=\mathrm{exp}(ikz)U(x),  \label{k}
\end{equation}%
cf. Eq. (\ref{U}), where real function $U(x)$ satisfies the equation
\begin{equation}
kU+\frac{1}{2}\left( -\frac{\partial ^{2}}{\partial x^{2}}\right) ^{\alpha
/2}U+V(x)U-gU^{3}=0.  \label{FNLSES}
\end{equation}%
Localized solutions, i.e., \textit{fractional solitons}, are characterized
by their power (alias norm, cf. Eq. (\ref{N2D}),
\begin{equation}
P=\int_{-\infty }^{+\infty }U^{2}(x)dx.  \label{P}
\end{equation}

The stability of these states against small perturbations is investigated by
looking for solutions in the form of
\begin{equation}
\Psi =\mathrm{exp}(ikz)[U(x)+a(x)\mathrm{exp}(\lambda z)+b^{\ast }(x)\mathrm{%
exp}(\lambda ^{\ast }z)],  \label{PERB}
\end{equation}%
where $a(x)$ and $b^{\ast }(x)$ represent the perturbation, and $\lambda $,
which may be complex, is the instability growth rate. Substituting
expression (\ref{PERB}) in Eq. (\ref{FNLSE}), one derives linearized
equations for the small perturbations,
\begin{eqnarray}
i\lambda a &=&\frac{1}{2}\left( -\frac{\partial ^{2}}{\partial x^{2}}\right)
^{\alpha /2}a+ka+gU^{2}(x)(2a+b),  \notag \\
i\lambda b &=&-\frac{1}{2}\left( -\frac{\partial ^{2}}{\partial x^{2}}%
\right) ^{\alpha /2}b-kb-gU^{2}(x)(2b+a).  \label{LAS}
\end{eqnarray}%
The stationary solution (\ref{k}) is stable if all eigenvalues produced by
the numerical solution of Eq. (\ref{LAS}) have Re$(\lambda )=0$.

Straightforward analysis of Eq. (\ref{FNLSE}) reveals a scaling relation
between the soliton's power and propagation constant:%
\begin{equation}
P(k,g)=P_{0}(\alpha )g^{-1}k^{1-1/\alpha },  \label{cubic}
\end{equation}%
with some coefficient $P_{0}(\alpha )$. Relation (\ref{cubic}) satisfies the
necessary stability condition, \textit{viz}., the celebrated \textit{%
Vakhitov-Kolokolov} (VK)\ criterion, $dP/dk>0$ \cite{VK,Berge} at $\alpha >1$%
, suggesting that the corresponding soliton family may be stable. The case
of $\alpha =1$, which corresponds to $dP/dk=0$ in Eq. (\ref{cubic}), implies
the occurrence of the \textit{critical collapse}, which makes all solitons
unstable, similar to the family of \textit{Townes solitons} produced by the
2D cubic nonlinear Schr\"{o}dinger equation with the normal (non-fractional)
diffraction ($\alpha =2$) \cite{Townes,Berge,book}. For $\alpha <1$,
relation (\ref{cubic}) contradicts the VK criterion, yielding $dP/dk<0$,
which implies strong instability of the solitons, driven by the \textit{%
supercritical collapse} \cite{Berge,book}.

Similar to what is outlined above for the system of FGPEs (\ref{+-}),
approximate fractal-soliton solutions of FNLSE (\ref{FNLSES}) can be sought
for by means of VA \cite{Chen,Frac14}. To this end, the Lagrangian of Eq. (%
\ref{FNLSES}) is used (cf. Eq. (\ref{L2D})),%
\begin{equation}
L=\frac{k}{2}\int_{-\infty }^{+\infty }dxU^{2}(x)+\frac{1}{4\pi }%
\int_{0}^{+\infty }dpp^{\alpha }\int \int d\xi dx\cos \left( p(x-\xi
)\right) U(x)U(\xi )-\frac{g}{4}\int_{-\infty }^{+\infty }dxU^{4}(x).
\label{Lagr}
\end{equation}%
The variational \textit{ansatz} for the solitons can be adopted as the usual
Gaussian with width $W$ and amplitude $A$ (cf. Eq. (\ref{ans2D})),%
\begin{equation}
U(x)=A\exp \left( -\frac{x^{2}}{2W^{2}}\right) ,  \label{ans}
\end{equation}%
whose power (\ref{P}) is
\begin{equation}
P=\sqrt{\pi }A^{2}W.  \label{P 2D}
\end{equation}%
The substitution of ansatz (\ref{ans}) in Lagrangian (\ref{Lagr}) and
integration yields the effective Lagrangian, in which the amplitude is
eliminated in\ favor of the power by means of Eq. (\ref{P 2D}):
\begin{equation}
L_{\mathrm{eff}}=-\frac{\mu }{2}P+\frac{\Gamma \left( \left( 1+\alpha
\right) /2\right) }{4\sqrt{\pi }}\frac{P}{W^{\alpha }}-\frac{g}{4\sqrt{2\pi }%
}\frac{P^{2}}{W},  \label{Leff}
\end{equation}%
cf. Eq. (\ref{va1}).

Typical examples of the fractional solitons with the parameters predicted by
the Euler-Lagrange equations, $\partial L_{\mathrm{eff}}/\partial P=\partial
L_{\mathrm{eff}}/\partial W=0$, and their comparison to numerical solutions
obtained for the same values of $\alpha $ and $k$ are displayed in Fig. \ref%
{Fig3}. Further, the characteristic of the soliton family in the form of the
$P(k)$ dependence, as predicted by the VA and produced by the numerical
solution, is plotted in Fig. \ref{Fig4}. These results demonstrate that VA
provides reasonable accuracy.
\begin{figure}[tbp]
\begin{center}
\includegraphics[width=0.76\columnwidth]{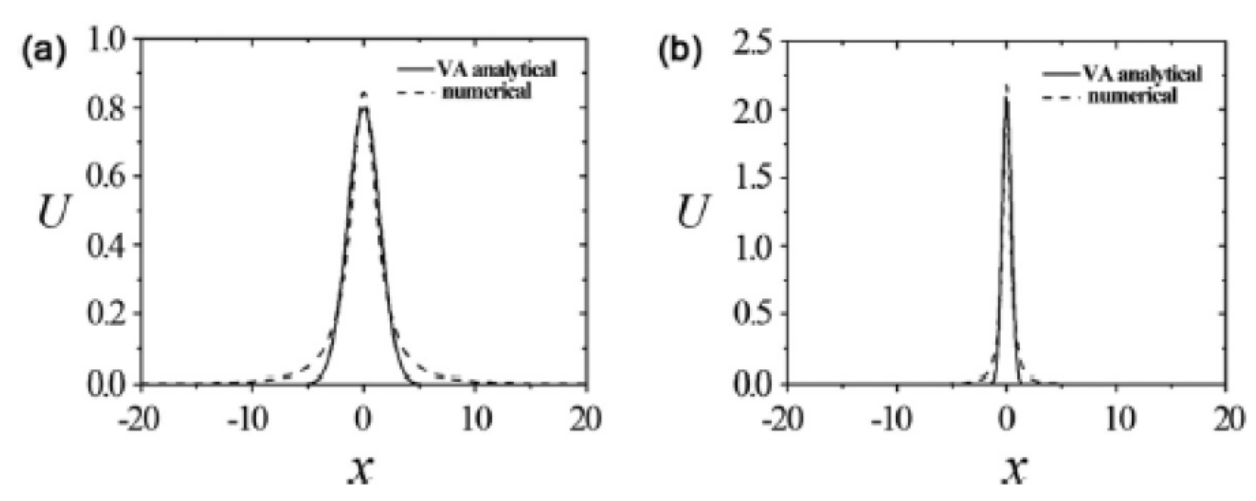}
\end{center}
\caption{Fractional solitons with $k=0.3$ (a) and $2.0$ (b), as predicted by
the VA based on ansatz \ (\protect\ref{ans}), and their counterparts
provided by the numerical solution of Eq. (\protect\ref{FNLSES}), for $%
\protect\alpha =1.5$ and $g=1$.
Reproduced with permission from
Y. Qiu, B. A. Malomed, D. Mihalache, X. Zhu, L. Zhang, and
Y. He, Soliton dynamics in a fractional complex Ginzburg-Landau model, Chaos
Solitons Fract. \textbf{131}, 109471 (2020) [see Ref.
\cite{Frac14}]. Copyright 2020, Elsevier.}
\label{Fig3}
\end{figure}
\begin{figure}[tbp]
\begin{center}
\includegraphics[width=0.42\columnwidth]{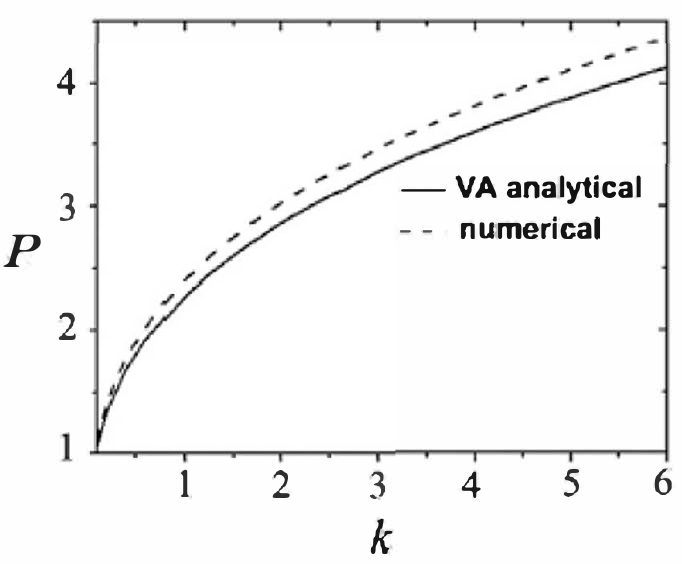}
\end{center}
\caption{Power $P$ of the family of the fractional solitons vs. the
propagation constant $k$, as predicted by the VA based on ansatz \ (\protect
\ref{ans}), and provided by the numerical solution of Eq. (\protect\ref%
{FNLSES}), for $\protect\alpha =1.5$ and $g=1$.
Reproduced with permission from
Y. Qiu, B. A. Malomed, D. Mihalache, X. Zhu, L. Zhang, and
Y. He, Soliton dynamics in a fractional complex Ginzburg-Landau model, Chaos
Solitons Fract. \textbf{131}, 109471 (2020) [see Ref.
\cite{Frac14}]. Copyright 2020, Elsevier.}
\label{Fig4}
\end{figure}

An important result of the VA is the prediction of the fixed power for the
\textit{quasi-Townes} solitons at $\alpha =1$, for which, as said above, the
critical collapse takes place \cite{Frac14}:%
\begin{equation}
\left( P_{\mathrm{Townes}}\right) _{\mathrm{VA}}=\sqrt{2}\approx 1.41.
\label{cubic-VA}
\end{equation}%
The numerically found counterpart of the VA prediction (\ref{cubic-VA}) is%
\begin{equation}
\left( N_{\mathrm{Townes}}\right) _{\mathrm{num}}\approx 1.23.
\label{cubic-num}
\end{equation}

Numerical investigation of the stability of the fractional-soliton family
demonstrates that some of them may be weakly unstable at values of LI which
are relatively far from $\alpha =2$. The instability does not destroy the
solitons, leading to the onset of small-amplitude intrinsic oscillations in
them \cite{Liangwei}.

\subsubsection{Quadratic nonlinearity in 1D and 2D}

In the case of quadratic (rather than cubic) self-focusing nonlinearity in
the 1D space, the critical collapse takes place at $\alpha =1/2$, hence
stable solitons may exist in the interval of $1/2<\alpha \leq 2$ \cite%
{Liangwei}, which is broader than its counterpart (\ref{1-2}) in the case of
the cubic self-focusing. In particular, the quadratic term may appear in Eq.
(\ref{FGPE}) as $-\varepsilon |\psi |\psi $ with $\varepsilon >0$,
representing a correction to the 1D FGPE induced by quantum fluctuations
around the mean-field state \cite{PA}.

In optics, the well-known quadratic nonlinearity appears in the system of
coupled FNLEs for the second-harmonic-generation model:%
\begin{eqnarray}
i\frac{\partial \Psi _{1}}{\partial z} &=&\frac{1}{2}\left( -\frac{\partial
^{2}}{\partial x^{2}}\right) ^{\alpha /2}\Psi _{1}+\Psi _{1}^{\ast }\Psi
_{2},  \notag \\
&&  \label{uv} \\
2i\frac{\partial \Psi _{2}}{\partial z} &=&\frac{1}{2}\left( -\frac{\partial
^{2}}{\partial x^{2}}\right) ^{\alpha /2}\Phi \Psi _{2}+Q\Psi _{2}+\frac{1}{2%
}\Psi _{1}^{2},  \notag
\end{eqnarray}%
where $\Psi _{1}$ and $\Psi _{2}$ are amplitudes of the fundamental and
second harmonics, and real $Q$ is the mismatch parameter, whose value may be
fixed, by means of scaling, as$Q=$ $\pm 1$ or $0$.

\begin{figure}[tbp]
\begin{center}
\includegraphics[width=0.72\columnwidth]{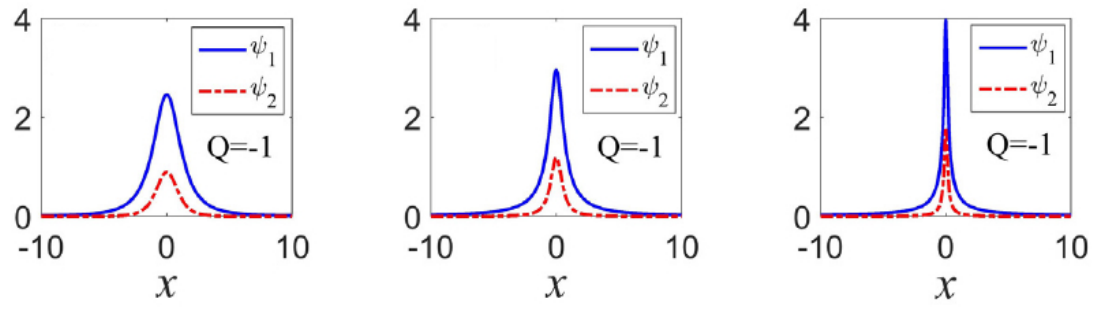}
\end{center}
\caption{Examples of stable solitons produced by Eq. (\protect\ref{12}) with
$\protect\alpha =1.5$, $1.0$, and $0.7$ (the left, central, and right
panels, respectively). In this case, $Q=-1$, and all the solitons correspond
to $\protect\beta =0.45$ in Eq. (\protect\ref{12}). The figure is borrowed
from Ref. \protect\cite{SHG}.
Reproduced from
P. Li, H. Sakaguchi, L. Zeng, X. Zhu, D. Mihalache, and B. A.
Malomed, Second-harmonic generation in the system with fractional
diffraction, Chaos, Solitons \& Fractals \textbf{173}, 113701 (2023)
[see Ref. \cite{SHG}] with the permission of Elsevier Publishing.}
\label{Fig5}
\end{figure}

Soliton solutions to Eq. (\ref{uv}), with real propagation constants $\beta $
and $2\beta $ of the fundamental-frequency and second-harmonic components,
respectively, are looked for as%
\begin{equation}
\Psi _{n}=\exp \left( in\beta z\right) \psi _{n}(x),~n=1,2,  \label{12}
\end{equation}%
where $\psi _{1,2}(x)$ are real localized profiles, see typical examples in
Fig. \ref{Fig5}. The existence and stability regions for solitons produced
by system (\ref{uv}) in its parameter space were identified in Ref. \cite%
{SHG}. In particular, the stability condition completely coincides with the
above-mentioned VK criterion, $dP/d\beta >0$, applied to the system's total
power, $P(\beta )=\sum_{n=1,2}\int_{-\infty }^{+\infty }n^{2}\psi
_{n}^{2}(x)dx$.

In the 2D space, the cubic self-focusing nonlinearity gives rise to the
critical collapse, as mentioned above, for $\alpha =2$, and to the
supercritical collapse at all values of $\alpha <2$, therefore all the 2D\
fractional solitons are unstable. 2D solitons, including ones with embedded
vorticity, may be stabilized in the model including the cubic self-focusing
and quintic defocusing. As shown in Ref. \cite{Frac6}, families of stable
solitons produced by the 2D FNLSE with the cubic-quintic nonlinearity and $%
\alpha <2$ are qualitatively similar to their counterparts which were
studied in detail in the 2D equation with the same nonlinearity and normal
(non-fractional) diffraction, corresponding to $\alpha =2$ \cite{Pego}.

On the other hand, the 2D version of the second-harmonic-generating system
based on Eq. (\ref{uv}) initiates the collapse only at $\alpha \leq 1$,
while 2D solitons may be stable in the above-mentioned intervals (\ref{1-2}%
). However, such solutions have not been investigated, as yet.

\subsubsection{Fast moving modes in the fractional medium: reduction to the
non-fractional equation}

The definition of the Riesz derivative, given by Eq. (\ref{Riesz derivative}%
), breaks the commonly known Galilean invariance of Eq. (\ref{FNLSE}) with
the non-fractional diffraction ($\alpha =2$), in the absence of the external
potential ($V=0$). Nevertheless, the FNLSE (as well as the linear FSE) can
be essentially simplified for fast moving modes (in the spatial domain the
\textquotedblleft moving" modes are actually ones tilted in the $\left(
x,z\right) $ plane) \cite{review}. To this end, one can consider solutions
built as a product of a slowly varying amplitude $\psi (x)$ and a rapidly
oscillating continuous-waver (CW) carrier with large wavenumber $\mathcal{P}$%
:%
\begin{equation}
\Psi \left( x,z\right) =\psi (x,z)e^{i\mathcal{P}x},  \label{psi}
\end{equation}%
The substitution of ansatz (\ref{psi}) in the nonlocally defined Riesz
derivative in Eq. (\ref{Riesz derivative}) leads to a quasi-local expression
expanded in powers of small $1/\mathcal{P}$:
\begin{equation}
\left( -\frac{\partial ^{2}}{\partial x^{2}}\right) ^{\alpha /2}\left( \psi
(x)e^{iPx}\right) =e^{iPx}|\mathcal{P}|^{\alpha }\left[ \psi
+\sum_{n=1}^{\infty }(-i)^{n}\frac{\alpha \left( \alpha -1\right) ...\left(
\alpha -n+1\right) }{n!P^{n}}\frac{\partial ^{n}\psi }{\partial x^{n}}\right]
.  \label{quasi}
\end{equation}%
Next, the substitution of expansion (\ref{quasi}), truncated at a finite
order, $n=n_{\max }$, in Eq. (\ref{FNLSE}) produces the local nonlinear Schr%
\"{o}dinger equation with higher-order diffraction terms corresponding to $%
n\geq 3$ in Eq. (\ref{quasi}). As concerns the convective term $\sim
i\partial \psi /\partial x$, which corresponds to $n=1$ in Eq. (\ref{quasi}%
), it may be eliminated by rewriting the resulting equation in the
coordinate system moving (tilted) with the group velocity%
\begin{equation}
V_{\mathrm{gr}}=\left( \alpha /2\right) |\mathcal{P}|^{\alpha -1}\mathrm{sgn}%
(\mathcal{P}),  \label{Vgr}
\end{equation}%
i.e., replacing coordinate $x$ by $\tilde{x}\equiv x-V_{\mathrm{gr}}z$.
Lastly, the term with $n=2$ in expansion (\ref{quasi}) gives rise to the
usual second-order diffraction term, with coefficient $D_{2}=(1/2)\alpha
\left( \alpha -1\right) |\mathcal{P}|^{-\left( 2-\alpha \right) }$. Thus,
the resulting equation, truncated at $n_{\max }=2$, takes the form of the
usual (non-fractional) nonlinear Schr\"{o}dinger equation,%
\begin{equation}
i\frac{\partial \tilde{\psi}}{\partial z}=-\frac{1}{2}D_{2}\frac{\partial
^{2}\tilde{\psi}}{\partial \tilde{x}^{2}}-g\left\vert \tilde{\psi}%
\right\vert ^{2}\tilde{\psi},  \label{psitilde}
\end{equation}%
which gives rise to the commonly known solutions, such as solitons.

\subsection{Systems of fractional nonlinear Schr\"{o}dinger equations with
the cubic nonlinearity}

\subsubsection{Fractional domain walls}

Extension of the fractional model for copropagating optical waves with
orthogonal polarizations or different carrier wavelength is a natural
possibility, which was addressed in Ref. \cite{Kumar} for the case of the
\emph{self-defocusing} cubic nonlinearity. The corresponding 1D system of
coupled FNLSEs for wave amplitudes $\Psi $ and $\Phi $ is%
\begin{eqnarray}
i\frac{\partial \Psi }{\partial z} &=&\frac{1}{2}\left( -\frac{\partial ^{2}%
}{\partial x^{2}}\right) ^{\alpha /2}\Psi +(|\Psi |^{2}+\beta |\Phi
|^{2})\Psi -\lambda \Phi ,  \notag \\
&&  \label{PsiPhi} \\
i\frac{\partial \Phi }{\partial z} &=&\frac{1}{2}\left( -\frac{\partial ^{2}%
}{\partial x^{2}}\right) ^{\alpha /2}\Phi +(|\Phi |^{2}+\beta |\Psi
|^{2})v-\lambda \Psi ,  \notag
\end{eqnarray}%
where $\beta >0$ is the relative strength of the cross-phase-modulation
(XPM) interaction, while the coefficient of the self-phase modulation is
scaled to be $1$. In the case of orthogonal polarizations, natural XPM\
values are $\beta =2/3$ or $2$, for the linear and circular polarizations,
respectively. The copropagation of optical waves carried by different
wavelengths is governed by the same system (\ref{PsiPhi}) with $\beta =2$
\cite{KA}. Other positive values of $\beta $ are possible in photonic
crystals \cite{KA}.

In addition to XPM, the two components may be coupled in Eq. (\ref{PsiPhi})
by linear terms with real coefficient $\lambda $. The linear coupling
accounts for the twist of the waveguide in the case of the copropagation of
linear polarizations, or elliptic deformation of the waveguide in the case
of circular polarizations \cite{KA}.

As usual, stationary states with a real propagation constant $k$ are looked
for as $\left\{ \Psi ,\Phi \right\} =\exp \left( ikz\right) \left\{
U(x),V(x)\right\} $, where real real functions $U(x)$ and $V(x)$ are
solutions of the system%
\begin{eqnarray}
kU+\frac{1}{2}\left( -\frac{\partial ^{2}}{\partial x^{2}}\right) ^{\alpha
/2}U+(U^{2}+\beta V^{2})U-\lambda V &=&0,  \notag \\
&&  \label{UV} \\
kV+\frac{1}{2}\left( -\frac{\partial ^{2}}{\partial x^{2}}\right) ^{\alpha
/2}V+(V^{2}+\beta U^{2})V-\lambda U &=&0.  \notag
\end{eqnarray}

Natural patterns supported by Eq. (\ref{UV}) with $k<0$ are optical domain
walls (DWs) \cite{DWs,Haelterman,PLA}, which link different CW states at $%
x\rightarrow \pm \infty $, that are mirror images to each other:

\begin{equation}
\left\{
\begin{array}{c}
U(x\rightarrow \pm \infty ) \\
V(x\rightarrow \pm \infty )%
\end{array}%
\right\} =\frac{1}{\sqrt{2}}\left\{
\begin{array}{c}
\sqrt{-\frac{k}{2}+\frac{\lambda }{\beta -1}}\pm \sqrt{-\frac{k}{2}-\frac{%
\lambda }{\beta -1}} \\
\sqrt{-\frac{k}{2}+\frac{\lambda }{\beta -1}}\mp \sqrt{-\frac{k}{2}-\frac{%
\lambda }{\beta -1}}%
\end{array}%
\right\} ,  \label{CW+}
\end{equation}%
$\allowbreak $with equal total power densities: $U^{2}(x\rightarrow \pm
\infty )+V^{2}(x\rightarrow \pm \infty )=-k$. In particular, in the absence
of the linear coupling, $\lambda =0$, $U(x)$ and $V(x)$ vanish at $%
x\rightarrow +\infty $ and $-\infty $, respectively. The asymmetric CW
states (\ref{CW+}) exist under the condition of the \textit{immiscibility}
\cite{immisc} of the two components:%
\begin{equation}
\beta -1>\left( \beta -1\right) _{\mathrm{immisc}}\equiv 2\lambda /|k|.
\label{max}
\end{equation}

In the case of $\beta =3$ (which plays a special role in various systems,
making it possible to find exact DW solutions \cite{old,PLA}), substitution $%
\left\{ U(x),V(x)\right\} =(1/2)\left[ \sqrt{\lambda -k}\mp W(x)\right] $
reduces two equations (\ref{UV}) to a single one,%
\begin{equation}
(k+\lambda )W+\frac{1}{2}\left( -\frac{\partial ^{2}}{\partial x^{2}}\right)
^{\alpha /2}W+W^{3}=0.  \label{W}
\end{equation}%
In this case, the boundary conditions (\ref{CW+}) which determined the DW
solutions reduce to
\begin{equation}
\lim_{x\rightarrow \pm \infty }W(x)=\pm (1/2)\sqrt{-k-\lambda }.  \label{bc}
\end{equation}%
Note that it follows from Eq. (\ref{max}) with $\beta =3$ that $\sqrt{%
-k-\lambda }$ is real, hence the boundary conditions (\ref{bc}) make sense.

\begin{figure}[tbp]
\begin{center}
\includegraphics[width=0.42\columnwidth]{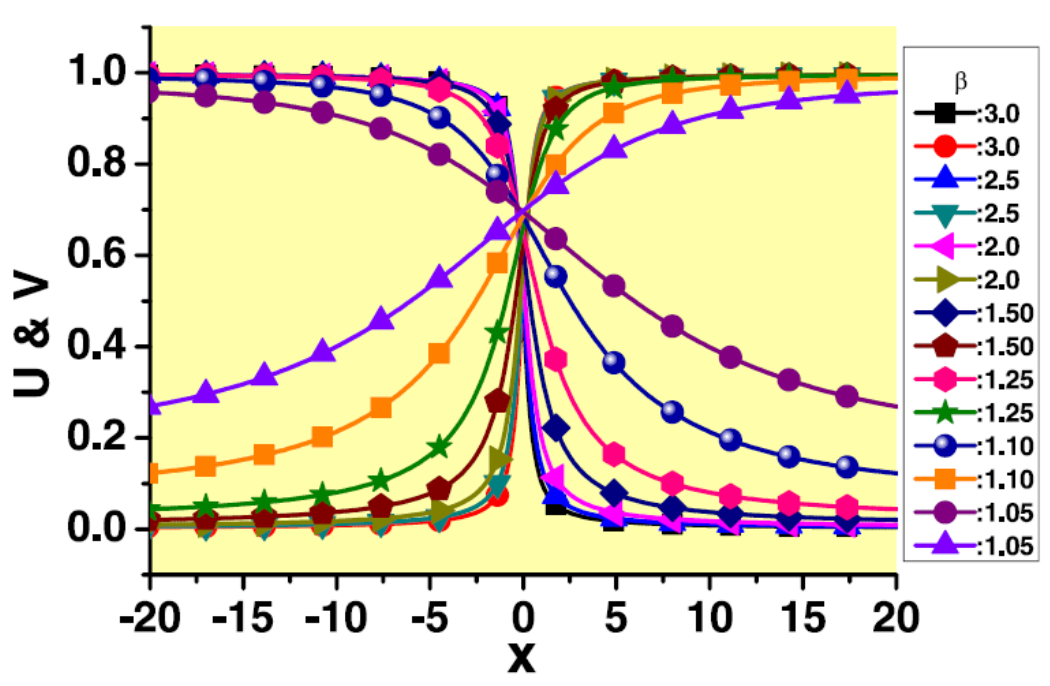}
\end{center}
\caption{A set of stable DW patterns with $k=-1$, produced by the numerical
solution of Eq. (\protect\ref{UV}) in the absence of the linear coupling ($%
\protect\lambda =0$), with LI $\protect\alpha =1$, and different values of
the XPM coefficient $\protect\beta $ indicated in the figure.
Reproduced from
S. Kumar, P. Li, and B. A. Malomed, Domain walls in
fractional media, Phys. Rev. E \textbf{106}, 054207 (2022)  [see Ref.
\cite{Kumar}] with the permission of AIP Publishing.}
\label{Fig6}
\end{figure}

The DW states produced by Eq. (\ref{PsiPhi}) are stable (in particular, the
CW background states at $x\rightarrow \pm \infty $ are not subject to
modulational instability). A set of numerically obtained DW solutions are
displayed in Fig. \ref{Fig6}.

A more general bimodal system, with different values of LI in its coupled
components, $\alpha _{1}$ and $\alpha _{2}$, was considered too \cite{Kumar}%
. It also gives rise to families of stable DW patterns, which, unlike those
displayed in Fig. \ref{Fig6}, are spatially asymmetric with respect to the
midpoint.

\subsubsection{Couplers and spontaneous symmetry breaking}

Another model which naturally gives rise to a system of coupled FNLSEs with
the fractional diffraction and cubic \emph{self-focusing} acting in each
equation, and linear coupling between the equations, introduces a dual-core
optical waveguide \cite{Frac18,Strunin}:%
\begin{eqnarray}
i\frac{\partial \Psi _{1}}{\partial z} &=&\frac{1}{2}\left( -\frac{\partial
^{2}}{\partial x^{2}}\right) ^{\alpha /2}\Psi _{1}-|\Psi _{1}|^{2}\Psi
_{1}-\Psi _{2},  \notag \\
&&  \label{coupler} \\
i\frac{\partial \Psi _{2}}{\partial z} &=&\frac{1}{2}\left( -\frac{\partial
^{2}}{\partial x^{2}}\right) ^{\alpha /2}\Psi _{2}-|\Psi _{2}|^{2}\Psi
_{2}-\Psi _{1},  \notag
\end{eqnarray}%
where the coefficient of the linear coupling is scaled to be $1$ (cf.
coefficient $\lambda $ in Eq. (\ref{PsiPhi})). As well as the single FNLSE (%
\ref{FNLSE}), system (\ref{coupler}) is meaningful in the interval of LI
index given by Eq. (\ref{1-2}), when the collapse does not take place.

Steady-state solutions to Eq. (\ref{coupler}) with real propagation constant
$k>0$ are looked for $\Psi _{1,2}=\exp \left( ikz\right) U_{1,2}(x)$, where
real functions $U_{1,2}(x)$ satisfy equations%
\begin{eqnarray}
&&kU_{1}+\frac{1}{2}\left( -\frac{d^{2}}{dx^{2}}\right) ^{\alpha
/2}U_{1}-U_{1}^{3}-U_{2},  \notag \\
&&  \label{couplerU} \\
&&kU_{2}+\frac{1}{2}\left( -\frac{d^{2}}{dx^{2}}\right) ^{\alpha
/2}U_{2}-U_{2}^{3}-U_{1},  \notag
\end{eqnarray}%
Two-component solitons produced by Eq. (\ref{coupler}) are characterized by
the total power,
\begin{equation}
P=\int_{-\infty }^{+\infty }\left[ U_{1}^{2}(x)+U_{2}^{2}(x)\right] dx.
\label{couplerP}
\end{equation}

In the case of the normal (non-fractional) diffraction, $\alpha =2$, a
fundamental property of the two-component solitons is \textit{spontaneous
symmetry breaking} (SSB), which destabilizes the obvious symmetric solitons,
with $U_{1}(x)=U_{2}(x)$, at a critical value of the soliton's power, $P=P_{%
\mathrm{cr}}$, and replaces the unstable symmetric modes by asymmetric ones,
with $U_{1}(x)\neq U_{2}(x)$. The asymmetry of the solitons is naturally
quantified by the normalized difference in powers of their components, i.e.,%
\begin{equation}
\Theta \equiv P^{-1}\int_{-\infty }^{+\infty }\left(
(U_{1}^{2}(x)-U_{2}^{2}(x)\right) dx.  \label{Theta}
\end{equation}

\begin{figure}[tbp]
\begin{center}
\includegraphics[width=0.72\columnwidth]{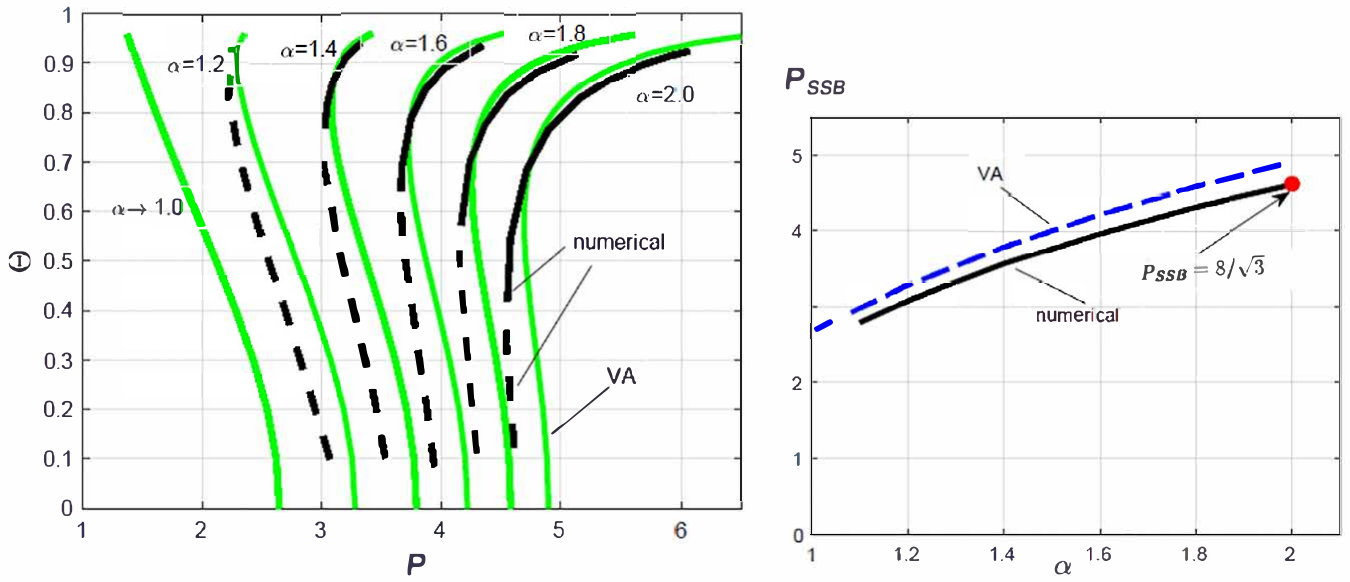}
\end{center}
\caption{Left: A set of dependences of the soliton's asymmetry parameter (%
\protect\ref{Theta}) on total power (\protect\ref{couplerP}), as predicted
by the VA through Eq. (\protect\ref{Th(N)}), and as produced by the
numerical solution of Eq. (\protect\ref{couplerU}), for different values of
LI $\protect\alpha $. Right: the value of the soliton's power at the SSB
bifurcation point vs. the LI, as predicted by the VA, and as found from the
numerical solution. The red circle at the end of the numerical curve
represents the known exact value (\ref{exact}) for $\protect\alpha %
=2 $ (the usual system with the non-fractional diffraction).
Reproduced from
D. V. Strunin and B. A. Malomed, Symmetry-breaking
transitions in quiescent and moving solitons in fractional couplers, Phys.
Rev. E \textbf{107}, 064203 (2023)
[see Ref. \cite{Strunin}] with the permission of AIP Publishing.}
\label{Fig7}
\end{figure}

The SSB phenomenology for solitons was studied in detail theoretically \cite%
{Wabnitz}-\cite{Peng}, and was recently demonstrated experimentally in
dual-core optical fibers \cite{Bugar} with the second-order (non-fractional)
GVD. A characteristic peculiarity of the SSB in this system is that the
transition from the symmetric solitons to asymmetric ones is performed
through a \textit{subcritical bifurcation} \cite{bif} (similar to a
symmetry-breaking phase transition of the first kind), which means that a
stable asymmetric soliton solution emerges at a value of $P$ which is
slightly smaller than $P_{\mathrm{cr}}$, i.e., in the \textit{subcritical}
range. In this case, the branch of the asymmetric soliton states, appearing
at $P=P_{\mathrm{cr}}$, originally moves in the backward direction, being
unstable (therefore, another name of this bifurcation is the \textit{backward%
} one). Then, this branch turns forward, getting stable, at the turning
point (see, e.g., the left panel of Fig. \ref{Fig7}).

The system of equations (\ref{couplerU}), with the fractional Riesz
derivative, defined as per Eq. (\ref{Riesz derivative}), can be derived from
the Lagrangian,%
\begin{equation}
L=\int_{-\infty }^{+\infty }\left[ \frac{k}{2}\left(
U_{1}^{2}+U_{2}^{2}\right) \right] dx+H,  \label{L}
\end{equation}%
where the Hamiltonian is%
\begin{gather}
H=\int_{-\infty }^{+\infty }\left\{ -\frac{1}{4}\left[
U_{1}^{4}(x)+U_{2}^{4}(x)\right] -U_{1}(x)U_{2}(x)\right\} dx  \notag \\
+\frac{1}{4\pi }\int_{0}^{+\infty }p^{\alpha }dp\int_{-\infty }^{+\infty
}dx\int_{-\infty }^{+\infty }dx^{\prime }\cos \left( p\left( x-x^{\prime
}\right) \right) \left[ U_{1}(x)U_{1}(x^{\prime })+U_{2}(x)U_{2}(x^{\prime })%
\right] ,  \label{H}
\end{gather}%
cf. Eq. (\ref{Lagr}). Using the Lagrangian, VA can be applied to the system.
An appropriate tractable ansatz, with width $W$, power $P$, and angle $\chi $
accounting for the norm distribution between the components, is%
\begin{equation}
U_{1}(x)=\sqrt{\frac{P}{2W}}\left( \cos \chi \right) \mathrm{sech}\left(
\frac{x}{W}\right) ,U_{2}(x)=\sqrt{\frac{P}{2W}}\left( \sin \chi \right)
\mathrm{sech}\left( \frac{x}{W}\right) .  \label{coupler ans}
\end{equation}%
In terms of this ansatz, the asymmetry parameter (\ref{Theta}) is%
\begin{equation}
\Theta _{\mathrm{VA}}=\cos \left( 2\chi \right) \equiv \sqrt{1-S^{2}}%
,S\equiv \sin \left( 2\chi \right) .  \label{ThetaVA}
\end{equation}

The calculation of Lagrangian (\ref{L}) with ansatz (\ref{coupler ans})
yields%
\begin{equation}
L_{\mathrm{eff}}=\frac{P}{2}\left[ k-\frac{P}{6W}\left( 1-\frac{1}{2}%
S^{2}\right) -S+\left( 1-2^{1-\alpha }\right) \Gamma (1+\alpha )\frac{\zeta
(\alpha )}{(\pi W)^{\alpha }}\right]  \label{Lcoupler}
\end{equation}%
(cf. Eq. (\ref{Leff})), where $\Gamma $ and $\zeta $ are the Gamma- and
zeta-functions. The effective Lagrangian (\ref{Lcoupler}) gives rise to the
Euler-Lagrange equations, $\partial L_{\mathrm{eff}}/\partial W=\partial L_{%
\mathrm{eff}}/\partial \left( \sin (2\chi \right) )=0$, which amount to
relation $W=PS/6$ and an equation for parameter $S$ (which is defined in Eq.
(\ref{ThetaVA})),%
\begin{equation}
S^{\alpha -1}\left( 1-\frac{S^{2}}{2}\right) =\frac{\alpha }{\pi ^{\alpha }}%
\left( 1-2^{1-\alpha }\right) \Gamma (1+\alpha )\zeta (\alpha )\left( \frac{6%
}{P}\right) ^{\alpha }.  \label{S}
\end{equation}

The SSB point, as which the family of the asymmetric solitons branches off
from the symmetric one, corresponds to setting $S=1$ (which represents the
symmetric soliton) in Eq. (\ref{S}), the result being%
\begin{equation}
\left( P_{\mathrm{SSB}}\right) (\alpha )=\frac{6}{\pi }\left[ 2\alpha \left(
1-2^{1-\alpha }\right) \Gamma (1+\alpha )\zeta (\alpha )\right] ^{1/\alpha }.
\label{N_SSB}
\end{equation}%
In the case of $\alpha =2$, Eq. (\ref{N_SSB}) reproduces the known result
produced by the VA for the non-fractional coupler, \textit{viz}., $\left( P_{%
\mathrm{SSB}}\right) (\alpha =2)=2\sqrt{6}\approx 4.899$ \cite{Pak,Peng}.
Note that the SSB point for the usual coupler is known in an exact form,
which can be found without the use of VA \cite{Wabnitz},%
\begin{equation}
\left( P_{\mathrm{SSB}}\right) _{\mathrm{exact}}(\alpha =2)=8/\sqrt{3}%
\approx 4.619,  \label{exact}
\end{equation}%
i.e., the relative error of the VA prediction is $\approx 5.7\%$. In the
opposite limit of $\alpha -1\rightarrow +0$, Eq. (\ref{N_SSB}) gives the
respective VA result as $\left( P_{\mathrm{SSB}}\right) \left( \alpha
\rightarrow 1\right) =(12/\pi )\ln 2\approx 2.648$.

Eventually, the VA predicts the basic characteristic of the system, \textit{%
viz}., the dependence of the asymmetry parameter (\ref{Theta}) on the power,
$\Theta (P)$, in an implicit form \cite{Strunin}:%
\begin{equation}
\left( 1-\Theta _{\mathrm{VA}}^{2}\right) ^{(\alpha -1)/2}\left( 1+\Theta _{%
\mathrm{VA}}^{2}\right) =\frac{2\alpha }{\pi ^{\alpha }}\left( 1-2^{1-\alpha
}\right) \Gamma (1+\alpha )\zeta (\alpha )\left( \frac{6}{P}\right) ^{\alpha
}.  \label{Th(N)}
\end{equation}%
In the limit of $\alpha \rightarrow 1$, this relation yields%
\begin{equation}
\Theta _{\mathrm{VA}}(P;\alpha \rightarrow 1)=\sqrt{\frac{12\ln 2}{\pi P}-1}.
\label{extreme}
\end{equation}

The VA predictions following from Eqs. (\ref{Th(N)}) and (\ref{N_SSB}) for
different values of LI $\alpha $ are plotted in Fig. \ref{Fig7}, along with
results produced by the numerical solution of Eq. (\ref{couplerU}). It is
observed that the subcritical character of the SSB bifurcation becomes more
pronounced with the decrease of $\alpha $. In the limit of $\alpha =1$, the
dependence is available solely in the VA form, given by Eq. (\ref{extreme}),
while the numerical solutions were obtained only for $\alpha \geq 1.2$ \cite%
{Strunin}. The VA-predicted $\Theta (P)$ dependence for $\alpha =1$, given
by Eq. (\ref{extreme}), takes the \textit{extreme subcritical form} (defined
as per Ref. (\cite{ECNU}), in which the branch of the asymmetric solitons,
going backward from the SSB point, never turns forward, always remaining
unstable). The dependence (\ref{extreme}) terminates at point $\Theta =1$,
i.e., at $P=(6/\pi )\ln 2\approx \allowbreak 1.\allowbreak 324$ (see the
left panel in Fig. \ref{Fig7}), as definition (\ref{Theta}) does not admit
values $|\Theta |>1$.

\begin{figure}[tbp]
\begin{center}
\includegraphics[width=0.50\columnwidth]{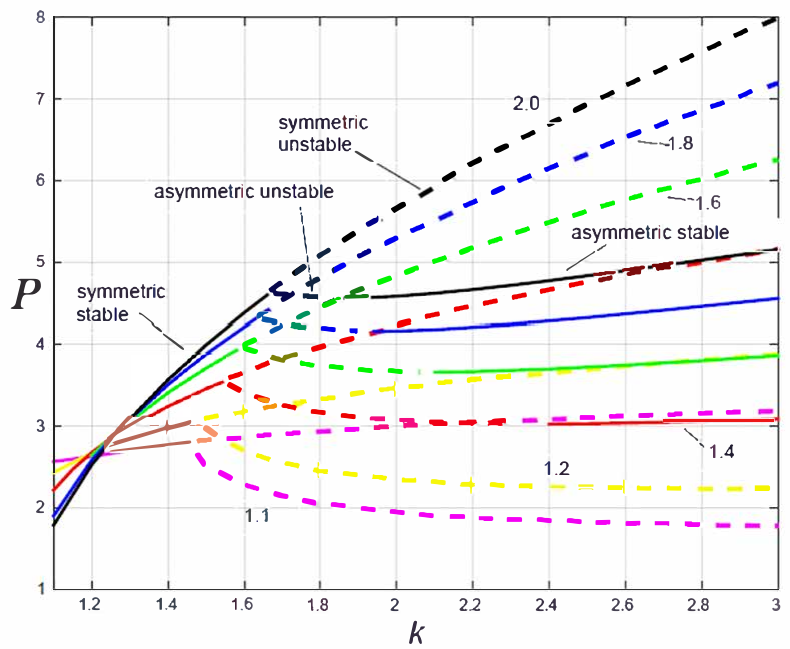}
\end{center}
\caption{Dependences between the total power, $P$, and the propagation
constant, $k$, for families of symmetric and asymmetric solitons produced by
the numerical solution of Eqs. (\protect\ref{coupler}) and (\protect\ref%
{couplerU}), for different fixed values of LI $\protect\alpha $. Solid and
dashed segments of the curves represent, respectively, stable and unstable
parts of the soliton families.
Reproduced from
D. V. Strunin and B. A. Malomed, Symmetry-breaking
transitions in quiescent and moving solitons in fractional couplers, Phys.
Rev. E \textbf{107}, 064203 (2023)
[see Ref. \cite{Strunin}] with the permission of AIP Publishing.}
\label{Fig8}
\end{figure}

In addition to that, Fig. \ref{Fig8} displays a set of numerically generated
dependences $P(k)$ for different fixed values of LI. The SSB points are
those separating stable and unstable segments of the symmetric-soliton
branches, from which ones representing unstable asymmetric solitons stem.
The stability boundaries, observed on the asymmetric branches for $\alpha
=2.0,1.8,1.6,$ and $1.4$, correspond to the above-mentioned turning points,
while the branches for $\alpha =1.2$ and $1.1$ do not reach those points,
therefore they seem completely unstable in Fig. \ref{Fig8}.

The analysis of the SSB phenomenology was also developed for moving (tilted)
two-component solitons (recall it is a nontrivial extension of the analysis
because the fractional diffraction breaks the system's Galilean invariance),
demonstrating that the symmetry-breaking bifurcation keeps its subcritical
character, and values of the power at the bifurcation point become lower
than for the quiescent solitons \cite{Strunin}. Collisions between moving
solitons were studied too, with a conclusion that the collisions with small
relative velocities lead to elastic rebound, which is followed by strongly
inelastic symmetry-breaking interactions at intermediate values of the
velocities, and, eventually, by restoration of the elasticity in collisions
between fast solitons \cite{Strunin}.

The analysis of SSB in solitons was also developed for the single-component
FNLSE (\ref{FNLSE}) with a symmetric double-well potential $V(x)$ \cite%
{Frac17}. On the contrary to what is shown here, in that case, the SSB
bifurcation is of the \textit{supercritical} (\textit{forward}) type (i.e.,
it represents a symmetry-breaking phase transition of the second type).
Furthermore, the same FNLSE with the self-defocusing nonlinearity ($g<0$ in
Eq. (\ref{FNLSE})) does not break the symmetry of the ground state trapped
by the double-well potential, but the increase of $|g|$ leads to spontaneous
breaking of the \textit{antisymmetry} of the first excited state trapped by
the same potential.

\section{The fractional group-velocity dispersion (GVD) in fiber cavities:
The fractional Schr\"{o}dinger equation in the temporal domain, and its
experimental realization}

\label{sec:4}

Thus far, no experimental realization of the effective fractional
diffraction for light beams in the spatial domain has been reported. Another
option for the implementation of the concept of the fractional propagation
in optics is to resort to the transmission of\ light pulses in the \textit{%
temporal domain}, i.e., in fiber-based setups, with an effective fractional
GVD. This option has been realized recently \cite{Shilong}, thus providing
the first experimental implementation of a fractional medium in any physical
setting.

Confining the consideration to the linear propagation regime, the effective
FSE for the optical amplitude $\Psi \left( \tau ,z\right) $ in the temporal
domain is written as
\begin{equation}
i\frac{{\partial \Psi }}{{\partial z}}=\left[ {\frac{D}{2}{{\left( {-\frac{{%
\partial ^{2}}}{{\partial {\tau ^{2}}}}}\right) }^{\alpha /2}}%
-\sum\limits_{k=2,3...}{\frac{{\beta _{k}}}{{k!}}{{\left( {i\frac{\partial }{%
{\partial \tau }}}\right) }^{k}}}}\right] \Psi ,  \label{E1}
\end{equation}%
where the evolutional variable is the propagation distance $z$, and the
effective coordinate is the \textit{reduced time} \cite{KA}, $\tau =t-z/V_{%
\mathrm{gr}}$, where $V_{\mathrm{gr}}$ is the group velocity of the carrier
wave. In this equation, the fractional GVD is determined by LI $\alpha $
(cf. Eq. (\ref{FSE})), $D$ is the corresponding coefficient, and the usual
derivatives represent the second- and higher-order GVD, for $k=2$ and $k\geq
3$, respectively, which act in the setup along with the fractional GVD.

The solution of Eq. (\ref{E1}) can be easily written for the Fourier
transform of the optical field,

\begin{equation}
\hat{\Psi}\left( \omega ,z\right) =\int_{-\infty }^{+\infty }\exp \left(
i\omega \tau \right) \Psi \left( \tau ,z\right) d\omega   \label{E2}
\end{equation}%
(cf. Eq. (\ref{Fourier})), as%
\begin{equation}
\hat{\Psi}(\omega ,z)=\exp \left( {-i\left( {\frac{D}{2}|\omega {|^{\alpha }}%
-\sum\limits_{k=2,3,...}{\frac{{\beta _{k}}}{{k!}}}{\omega ^{k}}}\right) z}%
\right) \hat{\Psi}_{\mathrm{input}}(\omega ),  \label{E3}
\end{equation}%
where $\hat{\Psi}_{\mathrm{input}}(\omega )$ is the Fourier transform of
initial condition, i.e., the optical signal coupled into the fiber at $z=0$.

\begin{figure}[tbp]
\begin{center}
\includegraphics[width=0.70\columnwidth]{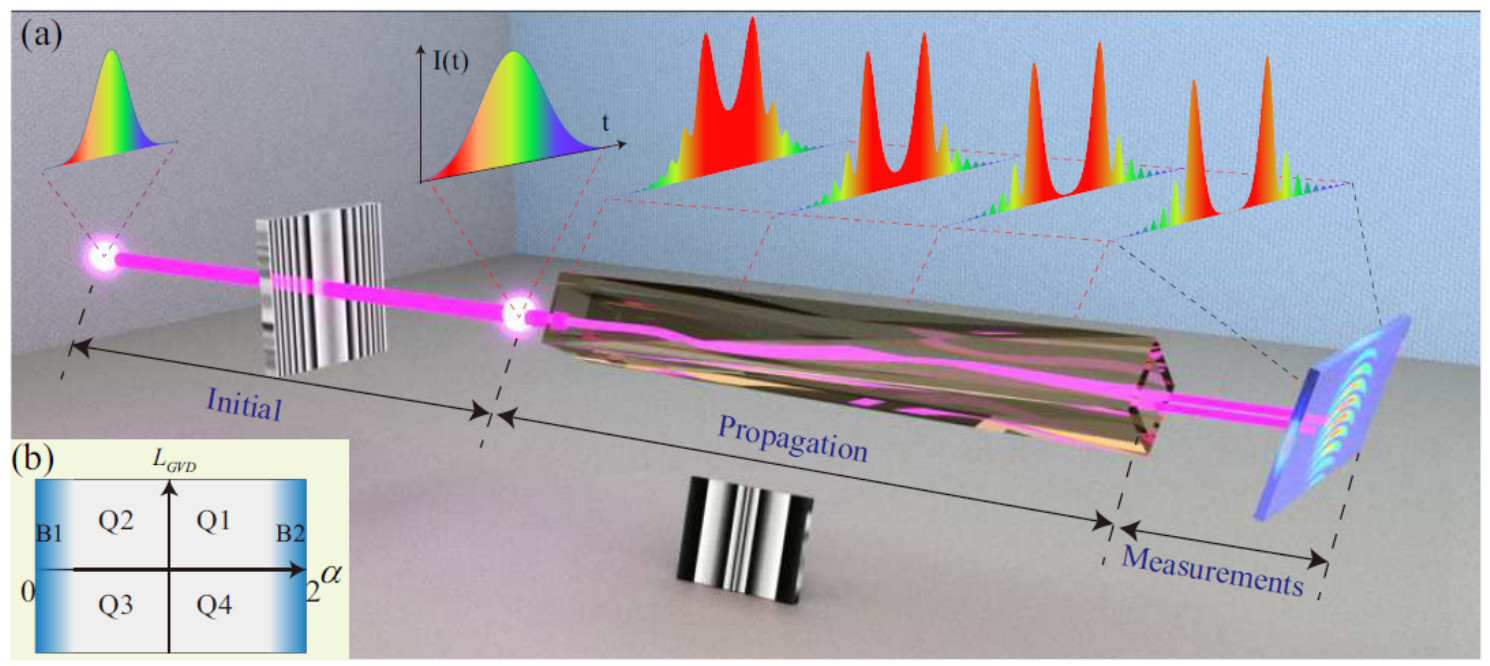}
\end{center}
\caption{(a) The setup elaborated in Ref. \protect\cite{Shilong} for the
realization of the fractional GVD\ in the fiber-laser cavity. The hologram
installed in the \textit{initial} section is employed to shape the input
pulse. In the \textit{propagation} section, the second hologram plays the
role of the phase mask (cf. Fig. \protect\ref{Fig2}), which imposes the
differential phase shift onto spectral components of the light signal, to
emulate the effect of the fractional GVD. This hologram is shown in the
rotated position, to display its structure. The initial temporal profile of
the pulse belonging to quadrant Q3 in (b), and its evolution in the \textit{%
propagation} section are shown in the top of the panel. (b) Four different
quadrants in the parameter plane ($L_{\mathrm{GVD}},\protect\alpha $). Q1
and Q2 correspond to the cases of $L_{\mathrm{GVD}}>0$ with $1\leq \protect%
\alpha \leq 2$ and $0\leq \protect\alpha \leq 1$, respectively. Q3 and Q4
are defined similarly for $L_{\mathrm{GVD}}<0$. Areas B1 and B2 designate
narrow strips with $\protect\alpha $ close to $0$ or $2$, respectively.
Reproduced with permission from
S. Liu, Y. Zhang, B. A. Malomed, and E. Karimi,
Experimental realisations of the fractional Schr\"{o}dinger equation in the
temporal domain, Nature Comm. \textbf{14}, 222 (2023)
[see Ref. \cite{Shilong}]. Copyright 2023, Springer-Nature.}
\label{Fig9}
\end{figure}

The fiber-cavity setup which was used for the experimental emulation of the
fractional GVD is displayed in Fig. \ref{Fig9}. It incorporates two
holograms, with the left one shaping the input pulse. The second hologram,
installed at the central position, was designed as the element similar to
the phase mask in the setup presented above in Fig. \ref{Fig2} (it is
displayed in the rotated form, to make its internal structure visible). It
imposes the differential phase shift onto spectral components of the light
signal, simulating the action of the fractional GVD as per Eq. (\ref{E3})
(cf. Eq. (\ref{Z})).

Experimental findings were compared with the theoretical result provided by
Eq. (\ref{E3}), keeping only the second-order GVD $\sim \beta $, in addition
to the fractional GVD \cite{Shilong}. Typical values of the dispersion
coefficients in Eq. (\ref{E3}) are
\begin{equation}
D=21\times 10^{-3}\text{ ps}^{\alpha }\text{/m},\beta _{2}=-21\times 10^{-3}~%
\text{ps}^{2}\text{/m,}  \label{Dbeta}
\end{equation}%
where $\beta _{2}<0$ implies that it represents the anomalous second-order
GVD \cite{KA}, although the case of the normal GVD, corresponding to $\beta
_{2}>0$, was considered too. The results were summarized as functions of two
control parameters, namely, LI $\alpha $ and effective GVD length $L_{%
\mathrm{GVD}}$ (which corresponds to $z$ in Eq. (\ref{E3})). The plane of
these parameters is schematically defined in the inset to Fig. \ref{Fig9},
where $\alpha $ varies from $0$ to $2$, while positive and negative values
of $L_{\mathrm{GVD}}$ correspond to the anomalous and normal second-order
GVD.

\begin{figure}[tbp]
\begin{center}
\includegraphics[width=0.92\columnwidth]{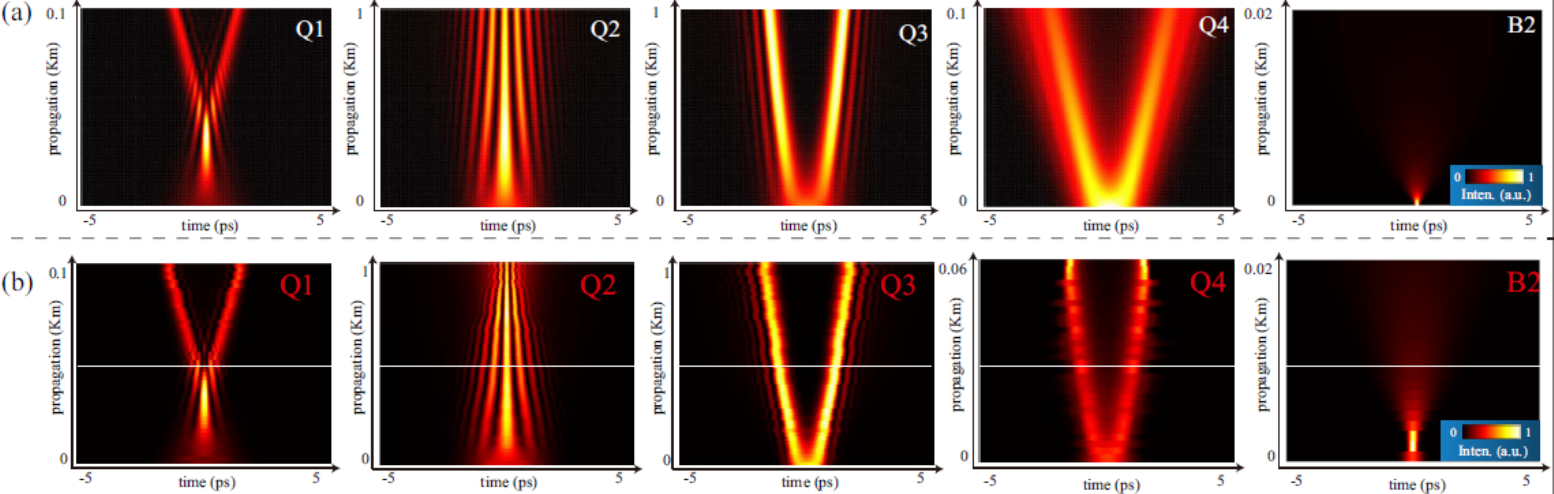}
\end{center}
\caption{Different outcomes of the propagation of input light pulses in
parameter regions Q1 -- Q4 and B2, defined as per Fig. \protect\ref{Fig9}%
(b), are displayed in the plane of the temporal coordinate $\protect\tau $
and propagation distance $z$, see Eq. (\protect\ref{E1}). The color code
represents local intensities of the pulses. Exact values of the parameters
corresponding to Q1 -- Q4 are fixed in Eq. (\protect\ref{control set}),
while B2 corresponds to $L_{\mathrm{GVD}}=0$ and $\protect\alpha =2$. In all
the cases, the dispersion parameters are fixed as per Eq. (\protect\ref%
{Dbeta}). Rows of panels (a) and (b) represent, severally, theoretical and
experimental results. Note the great difference in the scales of the
vertical axes ($z$) in panels Q2, Q3 ($z_{\max }=1$ km), Q1, Q4 ($z_{\max
}=0.1$ km) and B2 ($z_{\max }=0.02$ km).
Reproduced with permission from
S. Liu, Y. Zhang, B. A. Malomed, and E. Karimi,
Experimental realisations of the fractional Schr\"{o}dinger equation in the
temporal domain, Nature Comm. \textbf{14}, 222 (2023)
[see Ref. \cite{Shilong}]. Copyright 2023, Springer-Nature.}
\label{Fig10}
\end{figure}

Basic experimental and theoretical results are reported in Fig. \ref{Fig10}
for four characteristic cases, with parameters%
\begin{equation}
(L_{\mathrm{GVD}},\alpha )=(5,1.25),(5,0.25),(-5,0.25),(-5,1.25),
\label{control set}
\end{equation}%
that pertain, respectively, to quadrants Q1 -- Q4 in Fig. \ref{Fig9}(b). In
Q1, the setup demonstrates competition between the splitting effect of the
fractional GVD and the confining (anti-splitting) action of the anomalous
second-order GVD. The competition maintains balance over a short propagation
distance, $\simeq 40$ m. Then, the splitting effect becomes the dominant
one, leading to fission of the pulse in two secondary ones. In Q2, small $%
\alpha =0.25$ implies that the fractional GVD is weak. Then, the dominant
anomalous second-order GVD drives fragmentation of the input into a
multi-jet pattern, which, however, stays confined, following the confining
effect of the same GVD term observed at the initial stage of the evolution
in Q1. In panel Q3, the fractional GVD remains weak, with the same value of
LI, $\alpha =0.25$, as in Q2, while the second-order GVD flips its sign from
anomalous to normal, leading to the fast split of the pulse in two. Finally,
in quadrant Q4 the same normal second-order GVD as in Q3 cooperates with the
strong fractional-GVD term ($\alpha =1.25$), producing extremely fast
fission of the pulse in two fragments (note the difference by an order of
magnitude between scales on the vertical axes in panels Q3 and Q4). Another
difference between the relatively gentle splitting in Q3, driven by the
normal second-order GVD, and more violent splitting in Q4, driven jointly by
the second-order and fractional GVD terms, is that, in the latter case, the
emerging fragments are loosely bound pulses, unlike tightly bound ones in
Q3.

The dynamics displayed in Fig. \ref{Fig10} is reversible, therefore examples
of the splitting, such as ones exhibited in panels Q1 and Q3, imply the
possibility of \textit{fusion} of colliding pulses into a single one, if the
propagation distance $z$ varies in the opposite direction.

For the comparison's sake, panels B2 in Fig. \ref{Fig10} demonstrate that,
unlike the setup including the fractional GVD, the usual one, with $L_{%
\mathrm{GVD}}=0$ and $\alpha =2$ (see area B2 in the inset in Fig. \ref{Fig9}%
(b)), very quickly leads to full destruction of the input pulse (note that
largest propagation distance in panels B2 is $0.02$ km). Thus, the
fractional GVD makes the outcome of the light propagation in the fiber
cavity much more nontrivial.

\begin{figure}[tbp]
\begin{center}
\includegraphics[width=0.72\columnwidth]{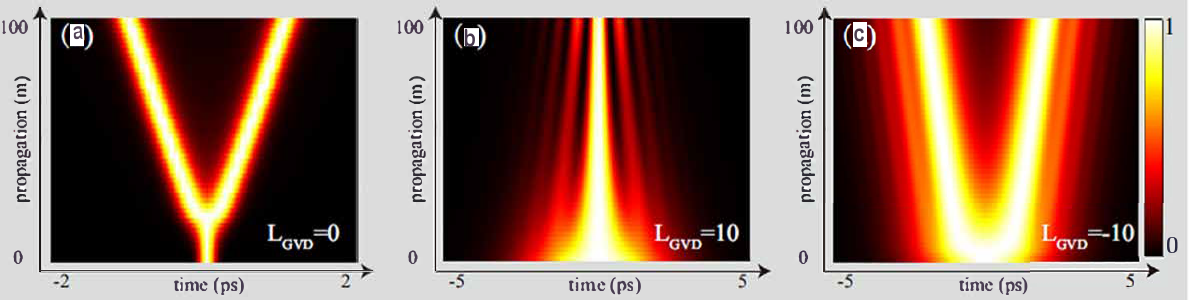}
\end{center}
\caption{Different outcomes of the propagation of the input pulses, as
predicted by simulations of the model based on Eq. (\protect\ref{E3}) with
the GVD parameters fixed as per Eq. (\protect\ref{Dbeta}), LI $\protect%
\alpha =1$, and values of $L_{\mathrm{GVD}}$ indicated in panels (a,b,c),
are displayed in the plane of $(\protect\tau ,z)$. The color code represents
local intensities.
Reproduced with permission from
S. Liu, Y. Zhang, B. A. Malomed, and E. Karimi,
Experimental realisations of the fractional Schr\"{o}dinger equation in the
temporal domain, Nature Comm. \textbf{14}, 222 (2023)
[see Ref. \cite{Shilong}]. Copyright 2023, Springer-Nature.}
\label{Fig11}
\end{figure}

A remarkable fact is close proximity between the experimental findings and
the corresponding theoretical results, which are presented, respectively, in
rows (b) and (a) of Fig. \ref{Fig10}. In addition to that, Fig. \ref{Fig11}
reports a set of theoretical results for the same dispersion parameters as
in Eq. (\ref{Dbeta}), but with another value of the LI, $\alpha =1$. In
particular, Fig. \ref{Fig11}(a) demonstrates that, under the action of
fractional-only GVD ($L_{\mathrm{GVD}}=0$), the evolution of the input is
generally similar to that in the case when the usual second-order GVD is
present too, but the fractional term is the dominant one (case Q1 in Fig. %
\ref{Fig10}): the input pulse quickly splits in a pair of secondary ones.
Such a setting with the purely fractional GVD was not realized
experimentally, as the usual GVD cannot be eliminated in the real setup.
Further, if strong anomalous or normal second-order GVD is included,
corresponding to $L_{\mathrm{GVD}}=+10$ and $-10$ in Figs. \ref{Fig11}(b)
and (c), respectively, the outcomes are also akin to what is shown for
similar cases in panels Q2 and Q4 of Fig. \ref{Fig10}: the fragmentation of
the input into a non-expanding multi-jet pattern in the former case, or
violent fission of the input in two loosely bound pulses in the latter case.

The experiments reported in Ref. \cite{Shilong} were also performed for the
input pulse with the phase including a cubic term, $\sim \tau ^{3}$. It is
well known that, in the framework of the usual linear Schr\"{o}dinger
equation with the second-order GVD, this term initiated the propagation of a
self-accelerating self-bending Airy wave \cite{Airy1,Airy2}. In the interval
of LI values $1\leq \alpha \leq 1.80$, the results reported in Ref. \cite%
{Shilong} (not shown here in detail) demonstrate splitting of the input into
a self-bending quasi-Airy wave and an additional one propagating along a
straight trajectory.\newline

All the experimental results demonstrating the implementation of the
effective fractional GVD, summarized above and reported in detail in Ref.
\cite{Shilong}, have been obtained in the linear regime. The interplay of
the strongly fractional GVD, with LI values $\alpha =1$ and $\alpha =0.2$,
and the fiber's self-focusing nonlinearity is a subject of a very recent
experimental work \cite{arXiv}. In particular, it demonstrates spectral
bifurcations of ultrashort optical pulses in the fiber-laser cavity, in the
form of spontaneous transitions between the pulses with single- and
multi-lobe structures.

\section{Conclusion}

\label{sec:5}

This aim of this paper is to present a concise summary of dynamical models
of 1D and 2D media with fractional diffraction or dispersion, both linear
and nonlinear ones. Two well-substantiated physical models of this type are
fractional quantum mechanics for particles which, at the classical level,
move by random L\'{e}vy flights \cite{Lask1,Lask2,GuoXu}, and the emulation
of the effective fractional diffraction in optical cavities \cite{Longhi}.
These models include the Riesz fractional derivatives \cite{Riesz} or the
respective 2D fractional Laplacian, which are defined as per Eqs. (\ref%
{Riesz derivative}) and (\ref{2D operator}). The Riesz derivative with LI (L%
\'{e}vy index) $\alpha $ amounts to the multiplication of the Fourier
transform of the underlying wave field, $\hat{\Psi}(p)$, by $|p|^{\alpha }$.
Equations (\ref{Riesz derivative}) and (\ref{2D operator}) demonstrate that
the Riesz derivatives and their 2D counterparts are represented, in the
explicit form, not by differential operators, but rather by nonlocal
integral ones.

The paper presents basic types of solitons produced by the nonlinear
fractional models in a brief form, without an objective to produce a
systematic review of the solitons in fractional systems. The absolute
majority of theoretical results for the fractional solitons have been
obtained by means of numerical methods. Nevertheless, the present review
includes some quasi-analytical results based on the VA (variational
approximation). Actually, the comparison with numerical findings
demonstrates that the VA works surprisingly well too in these relatively
complex nonlinear nonlocal models.

In addition to the theoretical results, the article includes a summary of
the recently reported first experimental realization of the fractional
optics. It is based on the fiber-cavity setup, in which the effective
fractional GVD (group-velocity dispersion) is implemented by means of a
specially designed phase plate (computer-generated hologram) \cite{Shilong}.

There are essential topics dealing with the dynamics of fractional media
which are not considered in this paper, which is not designed as a
comprehensive review. One of such topics is the dynamics of fractional
discrete systems. \cite{Tarasov,Molina,electrical}. Others, which have
attracted much interest recently (and may be relevant subjects for a more
extensive review), are dissipative solitons and self-trapped vortices in
models based on fractional complex Ginzburg-Landau equations \cite%
{Tarasov2,Juul,Arshed,Frac14}, as well as solitons in fractional
parity-time-symmetric systems \cite{Christodoulides,YaoLiu,Dong}.

There are many possibilities for further theoretical and, especially,
experimental developments in this area. In particular, on the theoretical
side, a challenging objective (which is mentioned in this paper) is to
derive equations of the FGPE (fractional Gross-Pitaevskii equation) type for
a BEC\ (Bose-Einstein condensate) of quantum particles which are
individually governed by the fractional Schr\"{o}dinger equation. On the
experimental side, obvious objectives are to realize the fractional
diffraction in spatial-domain optics, and to create the predicted fractional
solitons in such settings.

\section*{Acknowledgments}

I am grateful to Editors of this special issue of Chaos, celebrating the
80th birthday of David K. Campbell, for the invitation to submit a
contribution. I would like to thank my colleagues, with whom I had a chance
to collaborate on topics addressed in this brief review: Y. Cai, G. Dong, Y.
He, E. Karimi, S. Kumar, S. Liu, H. Long, X. Lu, P. Li, J. Li, T.
Mayteevarunyoo, D. Mihalache, Y. Qiu, H. Sakaguchi, D. Seletskiy, D.
Strunin, Q. Wang, J. Zeng, L. Zeng, L. Zhang, Y. Zhang, Q. Zhu, Y. Zhu, and
X. Zhu. My work on this topic was supported, in a part, by the Israel
Science Foundation through grant No. 1695/22.

\textbf{Acronyms}

\noindent 1D -- one-dimensional

\noindent \noindent 2D -- two-dimensional

\noindent BEC -- Bose-Einstein condensate

\noindent CQ -- cubic-quintic (nonlinearity)

\noindent CW -- continuous-wave (state)

\noindent DW -- domain wall

\noindent FGPE -- fractional Gross-Pitaevskii equation

\noindent FNLSE -- fractional nonlinear Schr\"{o}dinger equation

\noindent FSE -- fractional Schr\"{o}dinger equation

\noindent GVD -- group-velocity dispersion

\noindent LI -- L\'{e}vy index

\noindent SOC -- spin-orbit coupling

\noindent SSB -- spontaneous symmetry breaking

\noindent SV -- semi-vortex (a two-component 2D soliton, carrying
vorticities $0$ and $1$ in its components)

\noindent VA -- variational approximation

\noindent VK -- Vakhitov-Kolokolov (stability criterion)

\noindent XPM -- cross-phase-modulation

\end{document}